\pdfoutput=1
\documentclass[aps,twocolumn, floatfix, prx]{revtex4}
\usepackage{graphicx}
\DeclareGraphicsExtensions{.pdf}
\usepackage{amsmath,amssymb,bbold,bm,color}
\usepackage{float}
\usepackage{epstopdf}

\newcommand{\ba}{{\bm a}}
\newcommand{\bk}{{\bm k}}

\newcommand{\bp}{{\bm p}}
\newcommand{\br}{{\bm r}}
\newcommand{\bv}{{\bm v}}
\newcommand{\bR}{{\bm R}}
\newcommand{\bB}{{\bm B}}

\newcommand{\bG}{{\bm G}}

\newcommand{\bA}{{\bm A}}

\newcommand{\bj}{{\bm j}}
\newcommand{\bsig}{{\bm \sigma}}
\newcommand{\btau}{{\bm \tau}}
\newcommand{\bOmega}{{\bm \Omega}}

\newcommand{\cH}{{\cal H}}

\newcommand{\bee}{\begin{equation}}
\newcommand{\ee}{\end{equation}}

\def\sgn{\mathop{\rm sgn}\nolimits}

\begin{document}

\title{Electronic structure of topological superconductors in the
  presence of a vortex lattice
}

\author{Tianyu Liu}
\author{M. Franz}
\affiliation{Department of Physics and Astronomy, University of
British Columbia, Vancouver, BC, Canada V6T 1Z1}
\affiliation{Quantum Matter Institute, University of British Columbia, Vancouver BC, Canada V6T 1Z4}

\begin{abstract}
Certain types of topological superconductors and superfluids are known
to host protected Majorana zero modes in cores of Abrikosov
vortices. When such vortices are arranged in a dense periodic lattice
one expects zero modes from neighboring vortices to hybridize and form
dispersing bands. Understanding the structure of these bands is essential 
for the schemes that aim to employ the zero modes in quantum computation applications and in studies of their strongly interacting phases. 
We investigate here the band formation phenomenon in two
concrete models, describing a two dimensional $p_x+ip_y$
superconductor and a superconducting surface of a three-dimensional
strong topological insulator (Fu-Kane model), using a combination of
analytical and numerical techniques. We find that the physics of the
Majorana bands is well described by tight binding models of Majorana
fermions coupled to a static Z$_2$ gauge field with a non-trivial
gauge flux through each plaquette, in accord with expectations based
on very general arguments. In the case of the Fu-Kane model we also find
that, irrespective of the lattice geometry,  the Majorana band becomes completely flat at the so called neutrality point (chemical potential coincident with the Dirac point) where the model exhibits an extra chiral symmetry. In this limit the low energy physics will be dominated by four-fermion interaction terms which are permitted by symmetries and may arise from the Coulomb interaction between the constituent electron degrees of freedom.

\end{abstract}

\date{\today}

\maketitle

\section{Introduction}
Topological superconductors attract our attention in part because they
often host unpaired Majorana zero modes \cite{majorana1, wilczek1,
  franz1,alicea_rev,beenakker_rev,stanescu_rev,elliot_rev}. These in
turn exhibit a number of intriguing physical properties, including a
possibility to encode quantum information in a way that is robust to
environmental decoherence \cite{kitaev1} as well as to perform a
limited set of quantum gates that are topologically protected
\cite{read1,ivanov1,nayak1}. Thus far experimental evidence for
Majorana zero modes (MZMs) exists in quasi one-dimensional
systems including semiconductor quantum wires
\cite{mourik1,das1,deng1,rohkinson1,finck1,churchill1,Ramon} and wires composed of magnetic adatoms on a superconducting surface \cite{yazdani1,pawlak1}.
2D heterostructures made of strong topological insulators (STI)
and conventional superconductors (SC) are beginning to also show
promise \cite{koren1,sacepe1,qu1,williams1,cho1,elbaum1,xu0,xu1,xu2}. The
latter systems are predicted to host unpaired MZMs in cores of
Abrikosov vortices \cite{fu1} and the quantum information stored in
the zero mode subspace can be manipulated by performing adiabatic
exchanges -- ``braiding''-- of  the individual vortices. How exactly one performs such braiding operations, initializes the system and reads out the resulting quantum state remains largely an open question but is one of considerable interest.

A necessary first step towards the long term goal of storing and
manipulating quantum information in the Hilbert space spanned by MZMs
bound to vortex cores is to understand and characterize
these systems from the point of view of their electronic
structure. With this goal in mind we investigate here the electronic
structure of topological superconductors in the presence of Abrikosov
vortex lattices, paying particular attention to the fate of the MZMs
associated with individual vortices. Specifically, we study two
different models of electrons in a 2D topological superconductor in
the presence of a periodic vortex lattice. One describes a 2D spin
polarized $p_x+ip_y$ superconductor and the other a SC surface of a 3D
STI, also known as the Fu-Kane model \cite{fu1}. 
Although not explicitly studied here, we expect our results to apply to other realizations of 2D topological superconductors, such as those predicted to occur in heterostructures combining spin-orbit coupled semiconductors, ferromagnetic insulators and ordinary superconductors \cite{sau0,alicea0}.
 Because of the
interplay between the orbital effects of the applied magnetic field
$\bB$ that is necessary to establish the vortex lattice and the
spatially varying phase field $\theta(\br)$ of the SC order parameter
this turns out to be a problem of considerable subtlety and
complexity. A variant of this problem in the $p_x+ip_y$ superconductor
and an $s$-wave SC with Rashba spin-orbit coupling, but {\em
  neglecting} the applied magnetic field, has been studied in recent
works \cite{biswas1,kou1,kou2}. It is however well known
\cite{abrikosov1,tinkham1} that the vortex lattice is thermodynamically
unstable in the absence of $\bB$. (It is analogous to a system of
charged particles without an appropriate neutralizing background.) 

To understand the electronic structure of a physical vortex lattice the magnetic field must be properly included. We do this here using a technique of the singular gauge transformation \cite{tesanovic1} first developed in the context of high-$T_c$ cuprate superconductors with $d$-wave symmetry and subsequently applied to both $s$- and $p$-wave superconductors \cite{vafek1}. The chief advantage of this technique is that it treats the magnetic field $\bB$ and the SC phase field $\theta(\br)$ on an equal footing. Indeed we find results for low-energy Majorana modes that differ in several important aspects from the results of Refs. \cite{biswas1,kou1,kou2} obtained while neglecting the $\bB$ field. Most notably the flat Majorana bands predicted in Ref.\
\cite{biswas1} and the strong band structure anisotropies seen in
Ref.\ \cite{kou2} are not present in the full solution of the problem.
We conclude that magnetic field must be included in any calculation
that aims to correctly capture the physics of Majorana zero modes in a
realistic, thermodynamically stable vortex lattice. We note that
semiclassical treatment of a $p_x+ip_y$ superconductor has recently
been carried out (including the $\bB$ field) \cite{silaev1}  and showed results
consistent with our fully quantum mechanical
calculations.

Our main results can be summarized as follows. Denoting a Majorana
zero mode operator associated with a vortex core positioned at $\bR_j$
by $\gamma_j$ we find that in the presence of the vortex lattice the
physics of these modes is well described by a tight binding model of
the form
\begin{equation}\label{kin1}
{\cal H}_{\rm kin} =\sum_{i,j} \bar{t}_{ij}  \gamma_i\gamma_j.
\end{equation}
Here $\bar{t}_{ij}=t_{ij} s_{ij}$ can be decoposed into a real symmetric matrix $t_{ij}$ representing the hopping
strength while $s_{ij}=e^{i\phi_{ij}}=\pm i$ are Z$_2$ gauge
factors. The imaginary unity present in the latter is dictated by the Majorana commutation
relations 
\begin{equation}\label{can2}
\{\gamma_{i},\gamma_{j}\}=2\delta_{ij},
\ \ \ \gamma_{i}^\dagger=\gamma_{i},
\end{equation}
and the requirement that ${\cal H}_{\rm kin}$ be hermitian. The sign
ambiguity in $s_{ij} $ arises from the fact that one can perform a local
Z$_2$ gauge transformation $\gamma_j\to -\gamma_j$ without affecting
the zero mode commutation algebra  (\ref{can2}). A product of $s_{ij}$
factors along a closed path, however, represents a 
Z$_2$ gauge flux that is gauge invariant, and therefore in principle observable.  It is fixed by the microscopic Hamiltonian and can be thought of as analogous to the magnetic
flux expressed through Peierls factors in lattice models of charged
particles. Our main finding is that for the MZMs in the vortex lattice the Z$_2$ gauge flux through a general polygon formed by $n$ vortices  is given by the Grosfeld-Stern rule \cite{grosfeld1} 
\begin{equation}\label{kin2}
\sum_{\rm polygon}\phi_{ij}={\pi\over 2}(n-2),
\end{equation}
previously derived for MZMs in the Moore-Read fractional quantum Hall
state \cite{read2}, whose effective theory is analogous to the spin polarized $p_x+ip_y$ superconductor. 
Eq.\ (\ref{kin2}) indicates a non-zero Z$_2$ gauge flux ${\pi\over 2}$ and $\pi$
through an elementary triangular and square plaquette, respectively,
of the Majorana lattice. This in turn implies that the gapped phases
of the Hamiltonian (\ref{kin1}) are typically topologically
non-trivial with the occupied bands characterized by a non-zero Chern
number.  

We note that the  Majorana tight-binding model Eq.\ (\ref{kin1}) has been previously derived \cite{lahtinen0,lahtinen1} for vortex lattices present in the Kitaev spin model on the honeycomb lattice \cite{kitaev2}. It has been conjectured in these works that similar results should apply to other systems supporting localized Majorana mode arrays, but this conjecture has not yet been verified. Our work shows that the tight-binding model Eq.\ (\ref{kin1}) with the Z$_2$ gauge structure (\ref{kin2}) does apply to Abrikosov lattices in $p$-wave superconductors and the SC surfaces of topological insulators.

When the applied magnetic field is well below the upper
critical field $H_{c2}$ we furthermore find that the hopping amplitudes $t_{ij}$ are
significant only between the first and second nearest neighbors (nn). More
generally, $t_{ij}$ preserve all vortex lattice symmetries and exhibit an exponential decay $\sim
e^{-d_{ij}/\xi}$ with the distance $d_{ij}=|\bR_i-\bR_j|$ and $\xi$
the SC coherence length, superimposed on the RKKY-type oscillation with
a period close to the Fermi momentum $k_F$ of the underlying normal metal.
As already mentioned we find no sign of flat bands resulting from a subset of  
vanishing $t_{ij}$ predicted in Ref.\ \cite{biswas1} or anisotropies that break the underlying vortex lattice symmetry predicted in Ref.\ \cite{kou2}. These effects appear to be artifacts introduced by an approximation that neglects the magnetic field $\bB$. They may  be present in small clusters of vortices if such can be stabilized in the absence of $\bB$ but are not characteristic of a physical vortex lattice that retains its stability in the thermodynamic limit.

The results described above pertain to both the $p_x+ip_y$ superconductor and the Fu-Kane model. The latter shows an additional interesting feature when tuned to the neutrality point, reached when the chemical potential $\mu$ of the STI coincides with the Dirac point of the surface state. As noted previously \cite{teo1,galitsky1,galitsky2}, the model then exhibits an additional ``chiral'' symmetry which changes the topological classification of its zero modes from Z$_2$ to Z. Physically, this means that in the Fu-Kane model at neutrality MZM hybridization is prohibited and the Majorana band must remain flat irrespective of the geometry of the vortex lattice. We confirm by explicit numerical calculation that this is indeed the case. We also find that when slightly detuned from neutrality MZMs form a weakly dispersive band with a narrow bandwidth proportional to the chemical potential $\mu$ measured relative to the Dirac point. The Majorana flat band obtained by tuning a single parameter constitutes an interesting system because, just like in the fractional quantum Hall liquids \cite{tsui1,laughlin1}, the kinetic energy of the particles becomes quenched and the nature of the ground state is determined by interactions or disorder effects. If the sample is sufficiently clean so that disorder can be neglected and when interactions are present the system is inherently strongly correlated. Some consequences of these strong interactions in various 1D and 2D vortex lattice geometries have been explored in recent studies \cite{chiu1,chiu2,pikulin1,rahmani1,cobanera1,rahmani2}. Effects of disorder on the  Majorana tight-binding model Eq.\ (\ref{kin1}) have also been studied by several groups and interesting disorder-induced phases have been found \cite{lahtinen2,kraus1,vasuda1,lauman1,lauman2}.


\section{Majorana zero modes in vortex lattices}

Majorana zero modes associated with the individual vortices in $p_x+ip_y$ superconductor and the Fu-Kane model have been amply discussed in the literature \cite{elliot_rev,biswas1,galitsky1,galitsky2}. In this section we give a brief overview of some key results and then focus on the effect of the applied magnetic field on the collective behavior of the zero modes in such lattices. Using approximate analytical techniques we show how the low-energy Hamiltonian (\ref{kin1}) emerges in this setting and  give expressions for the overlap integrals $t_{ij}$ and the Z$_2$ gauge factors $s_{ij}$ valid in a physical vortex lattice that includes the magnetic field.

\subsection{Spin polarized $p_x+ip_y$ superconductor}

This is the simplest model of a 2D topological superconductor possibly
relevant to Sr$_2$RuO$_4$ \cite{kallin1}, the A phase of superfluid
$^3$He \cite{volovik1} and  the Moore-Read fractional quantum Hall
state \cite{read1}. The system is described by a second quantized
Hamiltonian
\begin{equation}\label{z0}
\cH=\int d^2r \hat{\Psi}^\dagger_\br H(\br) \hat{\Psi}_\br, \ \ \
\hat{\Psi}_\br= 
\begin{pmatrix}
c_\br \\
c^\dagger_\br
\end{pmatrix},
\end{equation}
where $c^\dagger_\br$ is a spinless fermion creation operator. 
The Bogoliubov-de Gennes (BdG) Hamiltonian has the form
\begin{equation}\label{z1}
H(\br)=
\begin{pmatrix}
\hat{h} & \hat{\Delta} \\
\hat{\Delta}^* & -\hat{h}^*
\end{pmatrix}
\end{equation}
where $\hat{h} = \bp^2/2m-\mu$ is the kinetic energy operator and
$\hat{\Delta}=k_F^{-1}\{\Delta(\br),\partial_x+i\partial_y\}$ is the
$p_x+ip_y$ pairing operator with $\Delta(\br)$ the SC gap
function. It respects the particle-hole symmetry generated by
$\Xi=\tau^xK$ where $\btau$ are Pauli matrices in the Nambu space and
$K$ denotes complex conjugation ($\Xi^2=1$). In addition it is
invariant under the global U(1) transformation $H\to e^{i\tau^z\chi}H
e^{-i\tau^z\chi}$ when accompanied by a phase shift $\Delta\to \Delta e^{-2i\chi}$

We are interested in the solutions of the BdG equation 
\begin{equation}\label{z2}
H\Phi(\br) =E\Phi(\br)
\end{equation}
in the presence of vortices in the SC order parameter
$\Delta(\br)=|\Delta(\br)|e^{i\theta(\br)}$ with  $\theta(\br)$ is the SC phase. In the presence of singly quantized vortices located at spatial positions $\{\bR_j\}$ we may write
\begin{equation}\label{z3}
\theta(\br)=\sum_k\varphi_k(\br), \ \ \ \varphi_k(\br)=\arg{(\br-\bR_k)}. 
\end{equation}
In addition $\Delta(\br)$ vanishes at the center of each
vortex and can be well approximated \cite{tinkham1} as $\Delta(\br)\simeq \Delta_0\prod_j
\tanh{(|\br-\bR_j|/\xi)}$. When the vortices are well separated so
that the smallest distance $d\gg\xi$ then we may look for the low
energy  solutions of the BdG equation (\ref{z2}) separately in the vicinity of each vortex. To this end we approximate the phase field near vortex $j$ as
\begin{equation}\label{z4}
\theta(\br)\simeq\varphi_j(\br) + \Theta_j
\end{equation}
where $\Theta_j=\sum_{k\neq j}\varphi_k(\bR_j)$ is the phase contributed by all other vortices in the system. Since by definition this contribution varies slowly near $\bR_j$ it is permissible to approximate it by a constant. For future reference we also note that in view of Eq.\ (\ref{z3}) we can write 
\begin{equation}\label{z4b}
\Theta_j=\theta(\bR_j)
\end{equation}
if we define $\theta(\bR_j)$ as being evaluated slightly to the right of the actual vortex position $\bR_j$, thus avoiding the singularity at the vortex center.

For a vortex $j$ the zero mode BdG wavefunction can thus be written as \cite{galitsky1}
\begin{equation}\label{z5}
\Phi_j(\br_j)=f(r_j)
\begin{pmatrix}
e^{i(\varphi_j+\Theta_j/2-\pi/4)}\\
e^{-i(\varphi_j+\Theta_j/2-\pi/4)}
\end{pmatrix},
\end{equation}
where $\br_j=\br-\bR_j$ and 
\begin{equation}\label{z6}
f(r)=\sqrt{k_F\over 2\pi\xi}J_1(k_F r)\exp{\left[-{1\over v_F}\int_0^r|\Delta(\br')|dr'\right]}.
\end{equation}
The quasiparticle operator $\gamma_j=\int d^2\br\Phi_j(\br_j)^\dagger\hat{\Psi}_\br$ associated with the zero mode has the property $\gamma_j^\dagger=\gamma_j$ and is therefore Majorana. At this level of approximation each vortex contains a single Majorana mode. These modes have zero energy, obey canonical commutation relations (\ref{can2}), and are separated from the rest of the spectrum by a minigap $\Delta_M\simeq \Delta_0^2/E_F$. Under adiabatic exchange vortices exhibit non-Abelian exchange statistics characteristic of the Ising anyons \cite{read1,ivanov1}.  

To understand the electronic structure of the zero modes beyond the
independent vortex approximation we must consider non-vanishing
overlaps between their wavefunctions (\ref{z5}). To leading order the resulting low-energy Hamiltonian takes the form of Eq.\ (\ref{kin1}) with 
\begin{equation}\label{z7}
\bar{t}_{ij}=\langle\Phi_i|H|\Phi_j\rangle.
\end{equation}
Refs.\ \cite{galitsky1,galitsky2} studied these overlap amplitudes
between two vortices in various limits and found characteristic
oscillatory RKKY-type behavior with an exponential decay as can be
expected on the basis of Eq.\ (\ref{z6}). Biswas \cite{biswas1}, in
addition pointed out a specific dependence on phase angles $\Theta_j$
of the form
\begin{equation}\label{z8}
\bar{t}_{ij}\propto \sin{\left(\Theta_i-\Theta_j\over 2\right)}.
\end{equation}
This follows directly from the spinor structure displayed in Eq.\
(\ref{z5}) and has important consequences for the collective behavior
of MZMs in situations with many vortices, such as in the Abrikosov lattice.

\subsection{Fu-Kane model}

Fu-Kane model describes a SC surface of a 3D STI and is defined by the
second quantized Hamiltonian (\ref{z0}) with a BdG Hamiltonian of the form
\begin{equation}\label{fk1}
H_{\rm FK}(\br)=
\begin{pmatrix}
\hat{h} & \hat{\Delta} \\
\hat{\Delta}^* & -\sigma^y\hat{h}^*\sigma^y
\end{pmatrix}
\end{equation}
where $\bsig$ are Pauli matrices acting in the physical spin space,
$\hat{h}=v\bp\cdot\bsig-\mu$ and $\hat{\Delta}={\rm
  diag}(\Delta(\br),\Delta(\br))$. The Hamiltonian acts on a
four-component spinor
$\hat\Psi_\br=(c_{\uparrow\br},c_{\downarrow\br},c^\dagger_{\downarrow\br},-c^\dagger_{\uparrow\br})^T$
in the combined spin and Nambu space. It respects the particle-hole symmetry generated by $\Xi=\tau^y\sigma^yK$  ($\Xi^2=1$) as well as the global U(1) symmetry defined below Eq.\ (\ref{z1}). 

The Hamiltonian (\ref{fk1}) is known to support unpaired Majorana zero modes in singly quantized vortices and antivortices \cite{fu1}. Their general properties have been explored in Ref.\ \cite{galitsky2}. Here we focus on the regime close to the neutrality point $\mu=0$ where $H_{\rm FK}$ exhibits an extra chiral symmetry generated by $\Pi=\tau^z\sigma^z$. As a result the structure of MZMs becomes particularly simple,
\begin{equation}\label{fk2}
\Phi_j(\br_j)=f_0(r_j)
\begin{pmatrix}
e^{i(\Theta_j/2-\pi/4)}\\
0\\
0\\
-e^{-i(\Theta_j/2-\pi/4)}
\end{pmatrix},
\end{equation}
with $f(r)=A \exp{[-{1\over v_F}\int_0^r|\Delta(\br')|dr']}$. The chiral symmetry has an important consequence that the overlap amplitudes $\bar{t}_{ij}$ between distinct MZMs exactly vanish at the neutrality point \cite{teo1,galitsky2}. When the symmetry is weakly broken by a small non-zero $\mu$ then the overlap becomes \cite{chiu1}
\begin{equation}\label{fk3}
\bar{t}_{ij}=i\mu F_{ij}\sin{\left(\Theta_i-\Theta_j\over 2\right)}
\end{equation}
with $F_{ij}=\int d^2r f_0(\br-\bR_i)f_0(\br-\bR_j)$.

\subsection{Inclusion of the magnetic field}

For the models discussed above to describe realistic vortex lattices magnetic field $\bB$ must be included in the theory. This is achieved by performing the minimal substitution $\bp\to \bp-{e\over c}\bA$ in Hamiltonians (\ref{z1}) and (\ref{fk1}) where $\bA$ is the vector potential such that $\bB=\nabla\times\bA$. Inclusion of the magnetic field preserves the discrete symmetries listed above but promotes the global U(1) symmetry to a gauge symmetry. Specifically, under the transformation $H(\br)\to e^{i\tau^z\chi(\br)}H(\br)
e^{-i\tau^z\chi(\br)}$ the Hamiltonians remain invariant provided that we transform the order parameter phase and the vector potential according to 
\begin{eqnarray}\label{g1}
\theta(\br)&\to& \theta(\br)-2\chi(\br), \\
\bA(\br) &\to& \bA(\br) -{\hbar c\over e}\nabla\chi(\br). \nonumber 
\end{eqnarray}
Here $\chi(\br)$ is an arbitrary smooth function.
It is now important to note that while the overlap amplitudes (\ref{z8}) and (\ref{fk3}) are properly invariant under the global U(1) symmetry, as written they are {\em not} invariant under the gauge transformation (\ref{g1}). Ref.\ \cite{chiu1} suggested that in the presence of the magnetic field the phase difference ${1\over 2}(\Theta_i-\Theta_j)$  be replaced by its gauge invariant counterpart 
\begin{equation}\label{g2}
\omega_{ij} =\int_{\bR_i}^{\bR_j}\left({1\over 2}\nabla\theta-{e\over
    \hbar c}\bA\right)\cdot d{\bf l},
\end{equation}
where the integral is taken along the straight line between $\bR_i$ and $\bR_j$.
In the remainder of this Section we shall justify this replacement in greater detail and we also evaluate the gauge invariant factors $\omega_{ij}$ in some specific situations of interest. 

In solving this problem we follow the classic procedure originally developed by Peierls \cite{peierls1} to include the magnetic field in the tight binding model for electrons moving in the ionic lattice. It relies on a key assumption that the applied magnetic field is sufficiently weak so that the vector potential $\bA(\br)$ can be replaced by a constant $\bA(\bR_j)$ for the purposes of obtaining the individual zero mode bound state $\Phi_j(\br)$. Such a constant vector potential can then be removed from the kinetic energy term in Eqs.\  (\ref{z1}) and (\ref{fk1}) by the gauge transformation (\ref{g1}) if $\chi(\br)$ is chosen such that
\begin{equation}\label{g3}
\nabla\chi_j(\br)={e\over \hbar c}\bA(\bR_j).
\end{equation}
In this gauge the zero mode is an eigenstate of the same Hamiltonian as in the absence of $\bB$ except the SC phase is now given by 
\begin{equation}\label{g4}
\tilde{\theta}(\br)=\sum_k\varphi_k(\br) -2\chi_j(\br).
\end{equation}
As before, near vortex $j$ we can separate the slowly varying part of the phase field and approximate it as
\begin{equation}\label{g5}
\tilde\theta(\br)\simeq\varphi_j(\br) + \tilde\Theta_j
\end{equation}
where $\tilde\Theta_j=\sum_{k\neq j}\varphi_k(\bR_j)-2\chi_j(\bR_j)$. The zero mode eigenstates are thus given, in this approximation, by Eqs.\ (\ref{z5}) and (\ref{fk2}) with $\Theta_j$ replaced by $\tilde\Theta_j$. The overlap integrals $\bar{t}_{ij}$ in the presence of the magnetic field can be computed using Eqs.\ (\ref{z8}) and (\ref{fk3}) with the same replacement for $\Theta_j$.

An important subtle point here is that Eq.\ (\ref{g3}) defines $\chi_j(\br)$ only up to an additive constant. The overlap amplitudes $\bar{t}_{ij}$ will have an invariant meaning only if this constant is chosen to be the same for all $j$ because then it  drops out of all differences  $\tilde\Theta_i-\tilde\Theta_j$. This condition is conveniently implemented by  making use of Eq.\ (\ref{g2}) in which the integrand is manifestly gauge invariant, corresponding to a consistent choice of the additive constant. Specifically, we conclude that in the presence of magnetic field Eq.\ (\ref{fk3}) is replaced by 
\begin{equation}\label{fk3b}
\bar{t}_{ij}=i\mu F_{ij}\sin{\omega_{ij}}
\end{equation}
with the gauge invariant phase difference defined in Eq.\ (\ref{g2}).
The integral itself is path dependent but, as argued by Peierls \cite{peierls1}, the straight line choice is most physical because for exponentially localized orbitals the actual tunneling path is predominantly along the straight line where the overlap wavefunction amplitude is maximal. The standard Peierls substitution based on this reasoning is known to provide an accurate description of itinerant electrons moving in ionic lattices subject to magnetic fields. We will demonstrate below, using extensive numerical simulations, that its generalization (\ref{g2}) to Majorana fermions in vortex lattices likewise provides a description that is both qualitatively and quantitatively accurate.

\subsection{Computation of the phase factors and the Z$_2$ gauge structure}

According to the previous subsection in the presence of the magnetic field the overlap integrals defined by Eqs.\ (\ref{z5}) and (\ref{fk2}) are to be calculated replacing ${1\over 2}(\Theta_i-\Theta_j)$ by $\omega_{ij}$ defined by Eq.\ (\ref{g2}). While the amplitude of $\bar{t}_{ij}$ depends strongly on various parameters of the model as well as on the distance $d$ between the vortices, the Z$_2$ gauge factors, which we define as  
\begin{equation}\label{x1}
s_{ij}=i\sgn{(\sin{\omega_{ij}})}=\pm i,
\end{equation}
are universal in that they depend only on the vortex lattice geometry. In the following we outline the general procedure for the evaluation of these gauge factors and we also find them explicitly for some simple lattice geometries.

We are mostly interested in physical situations when the SC forms a thin quasi-2D layer. In this case the effective penetration depth is given by the Pearl length $\lambda_{\rm eff} =2\lambda_L^2/h$ where $\lambda_L$ is the bulk penetration depth and $h$ the thickness of the SC film. In most cases we expect  $\lambda_L\gg h$ making $\lambda_{\rm eff}$ very large. This in turn means that the magnetic field can be taken as essentially constant in space whenever $d\lesssim \lambda_{\rm eff}$. 

To calculate the phase factors $\omega_{ij}$ defined in Eq.\ (\ref{g2}) it is useful to denote the integrand 
\begin{equation}\label{x2}
\bOmega={1\over 2}\left(\nabla\theta-{2e\over \hbar c}\bA\right)
\end{equation}
and recall that 
\begin{equation}\label{x3}
\nabla\times\bOmega=\pi\hat{z}\left[\sum_j\delta(\br-\bR_j)-{B\over\Phi_0^*}\right],
\end{equation}
where $\Phi_0^*=hc/2e$ is the SC flux quantum. $\bOmega$ can thus be thought of as a vector potential of a fictitious magnetic field that consists of $\delta$-function $\pi$ fluxes associated with vortex singularities on top of a neutralizing, almost uniform physical magnetic field contributing flux $-\pi$ per vortex. This picture will be useful for determining $\omega_{ij}$ in vortex lattices with high symmetry. One can evaluate $\bOmega$ by noting that it is related to the physical supercurrent $\bj_s$ through 
\begin{equation}\label{x4}
\bj_s=n_s{2\hbar e^*\over m^*}\bOmega,
\end{equation}
where  $n_s$ represents the superfluid density while $e^*=2e$ and $m^*$ are, respectively, the effective charge and mass of the Cooper pair. Combining Eqs.\ (\ref{x3},\ref{x4}) with the Amp\`{e}re's law $\nabla\times\bB=(4\pi/c)\bj_s$ one obtains the London equation for $\bB=\hat{z}B$ in the vortex lattice \cite{tinkham1},
\begin{equation}\label{x5}
B-\lambda_L^2\nabla^2B=\Phi_0^*\sum_j\delta(\br-\bR_j),
\end{equation}
where $\lambda_L^2=mc^2/4\pi{e^*}^2 n_s$ is the London penetration depth.
For a periodic lattice the equation can be solved by Fourier transforming, 
\begin{equation}\label{x6}
\bB(\br)=\Phi_0^*\hat{z}\sum_\bG{e^{i\bG\cdot\br}\over 1+\lambda_L^2G^2},
\end{equation}
where the sum extends over all reciprocal vectors $\bG$ of the vortex lattice. From the knowledge of $\bB$ one can reconstruct $\bOmega$ via Eq.\ (\ref{x3}) obtaining
\begin{equation}\label{x7}
\bOmega(\br)=\pi\sum_\bG{i\bG\times\hat{z}\over \lambda_L^{-2}+G^2}e^{i\bG\cdot\br}.
\end{equation}
The gauge invariant phase factors $\omega_{ij}$ can now be determined by a straightforward integration of $\bOmega(\br)$ in Eq.\ (\ref{g2}) followed by a numerical evaluation of the reciprocal lattice vector sums.

\begin{figure}[t]
\includegraphics[width = 8.0cm]{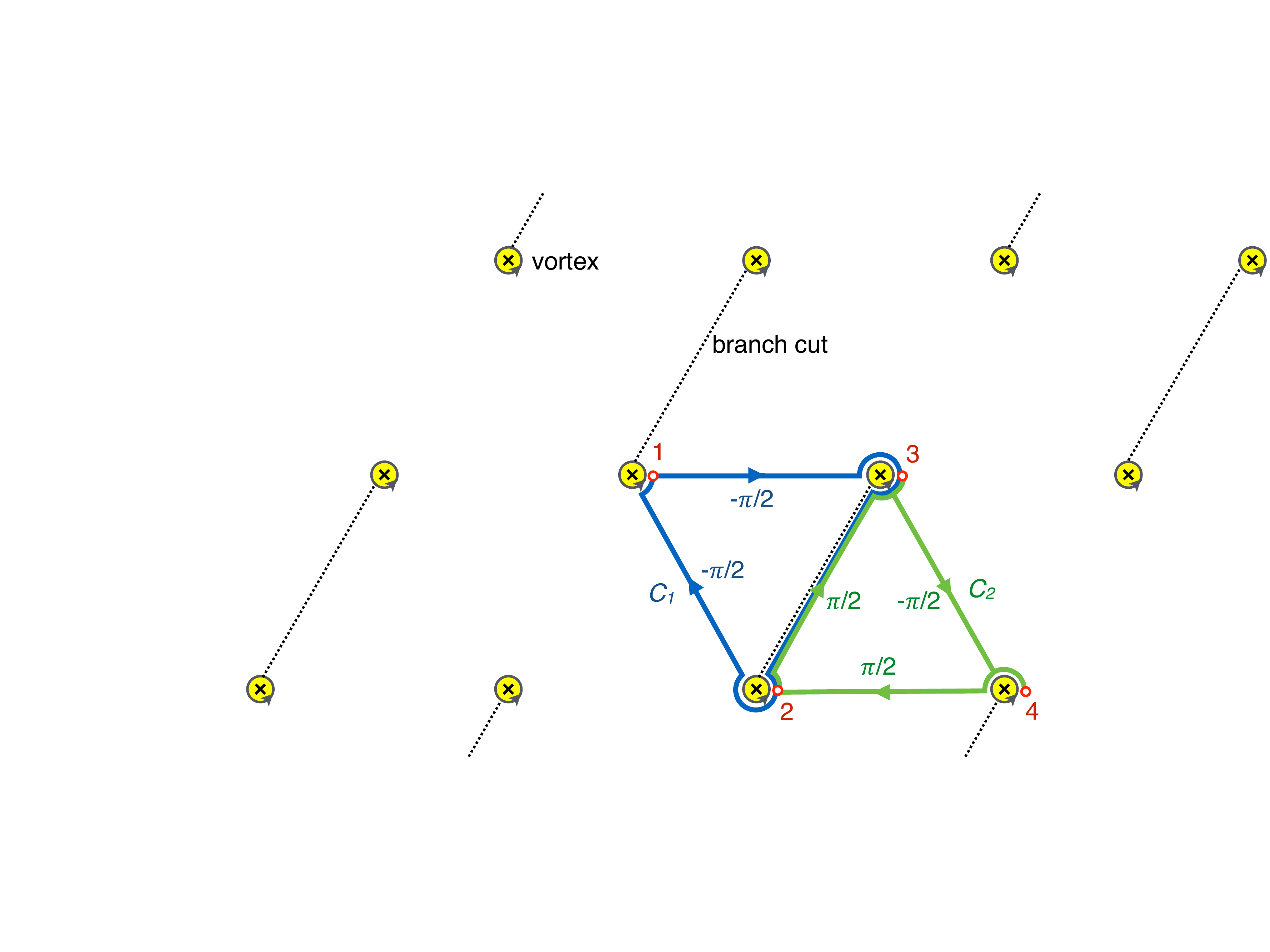}
\caption{Phase factors and branch cuts in a triangular vortex
  lattice. Oriented solid lines indicate integration paths between the
  reference points located just to the right of each each vortex
  center. These are used to evaluate the gauge invariant phase factors
  $\omega_{ij}$. Dashed lines represent a specific choice of the branch cuts discussed in the text.  
}\label{fig1}
\end{figure}
For a regular periodic vortex lattice with high symmetry, such as
the triangular lattice depicted in Fig.\ \ref{fig1}, it is possible to
determine the gauge invariant phase factors $\omega_{ij}$ without
resorting to detailed calculations. One can argue in two
stages. First, consider a closed
path  $C_1$ indicated in Fig.\ \ref{fig1}. It consists of straight line segments and circular segments. In the following we shall consider the latter to have infinitesimal radii. The integral $\oint_{C_1}\bOmega\cdot
d{\bf l}=\int(\nabla\times\bOmega)\cdot dS$ can be seen to equal
$-3\pi/2$ as the path encircles two vortices each contributing flux
$-\pi$ plus an area pierced by magnetic flux $\Phi_0^*/2$ contributing
$+\pi/2$ to the integral. It is also easy to see that the circular segments
of the path alone contribute the same
amount of flux $-3\pi/2$. This shows that the total contribution of the straight line
segments must be zero. On symmetry grounds we furthermore expect each straight
segment to give the same contribution which must therefore be
zero. The same conclusion can be reached by considering another path,
such as $C_2$, indicating that this result is consistent. We
thus arrive at a simple recipe for finding $\omega_{ij}$: straight
line segments contribute zero while the circular segments contribute
a phase $\pm\alpha/2$ where $\alpha$ is the angular length of the
segment and the sign depends on the sense of rotation with the $+$
sign taken for counterclockwise rotation.  

Second, we must attend to the branch cuts. To motivate this consider
the path $1\to 3$ in Fig.\ \ref{fig1}. According to the above recipe we have   $\omega_{13}=-{\pi/2}$. However, had we avoided the singularity associated with
vortex 3 from below the result  would have been
$+\pi/2$. More generally, taking the opposite path around the vortex
can be seen to change $\omega_{ij}\to\omega_{ij}\pm \pi$.  This ambiguity has to do
with the fact that $\omega_{ij}$ as defined in Eq.\ (\ref{x2}) has an
overall factor of ${1\over 2}$ in front of the phase gradient which
makes $\sin{\omega_{ij}}$ non-single valued in the presence of
  vortices. Importantly, this non-single valuedness underlies the Z$_2$ gauge
  structure present in the tight binding model (\ref{kin1}). In order
  to produce a consistent low-energy theory for the MZMs we must
  specify $\omega_{ij}$ in a globally unique fashion. This is achieved
  by defining branch cuts, emanating one from each vortex, along which
  $\bOmega$ varies discontinuously. It is most
  convenient to choose the branch cuts such that they terminate in a nearby
  vortex. One such choice of the branch cuts is illustrated by dashed
  lines in Fig.\ \ref{fig1}. Integration paths between points
  $\bR_i$ and $\bR_j$ chosen so as not to  cross any branch cuts then
  furnish a globally unique definition of $\omega_{ij}$ which
  corresponds to a particular choice of the Z$_2$ gauge. A different
  choice of the branch cuts corresponds to a different  Z$_2$ gauge
  but leaves all physical observables invariant. Factors $\pm\pi/2$
  indicated in Fig.\ \ref{fig1} have been obtained according to this
  prescription and can be seen to obey the Grosfeld-Stern rule Eq.\ (\ref{kin2}). A similar analysis can be performed for the square vortex lattice \cite{chiu1} and leads to the same conclusion.

\subsection{Tight binding dispersions for Majorana bands}
\begin{figure}[t]
\includegraphics[width = 8.0cm]{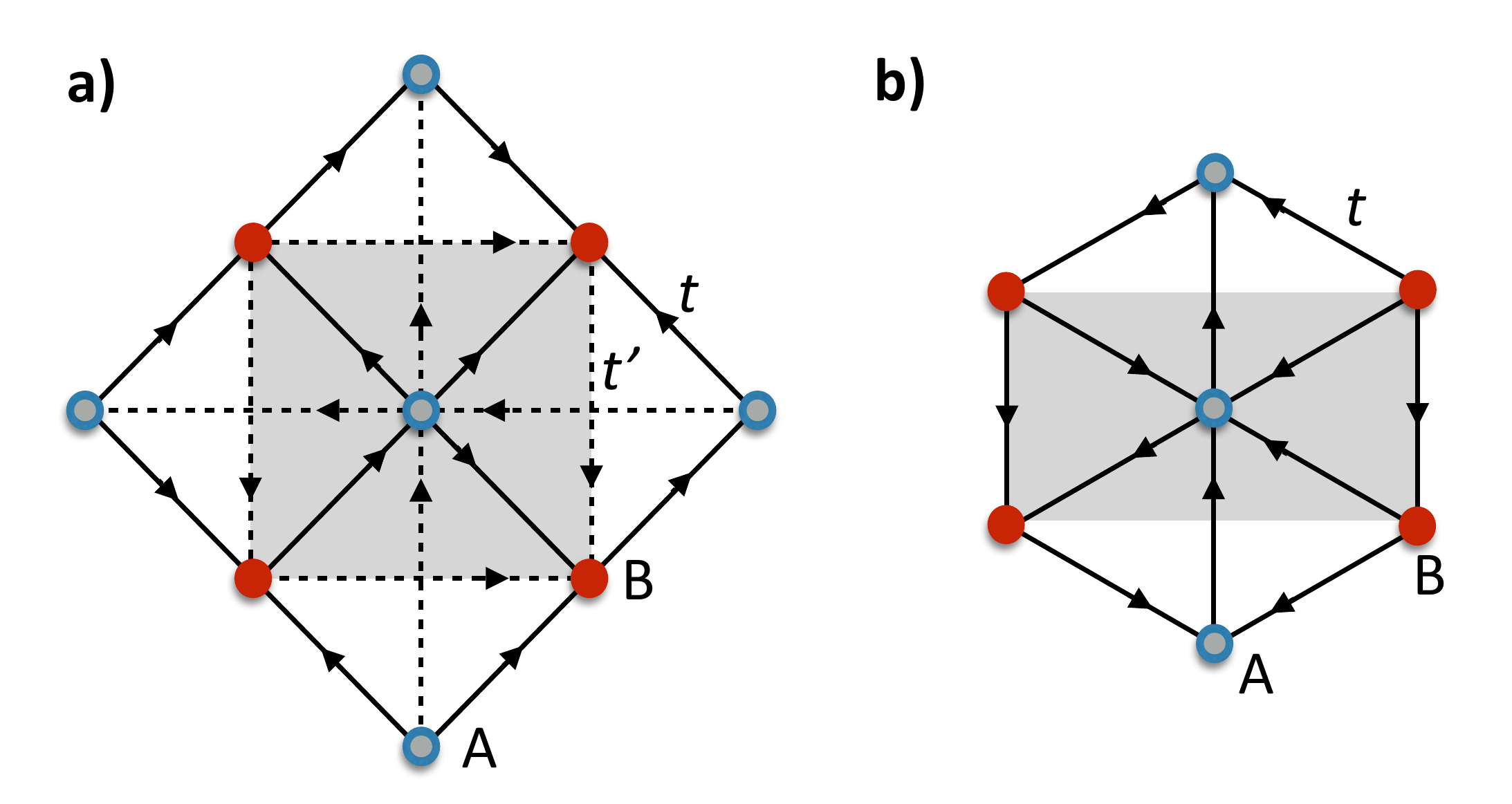}
\caption{Vortex lattice geometries: a) square and b) triangular. Two-vortex unit cell is shaded. The arrows specify the Z$_2$ gauge factors for the MZM tight binding models and satisfy the Grosfeld-Stern rule Eq. (\ref{kin2}). Hopping in the direction of the arrow incurs a phase factor of $+i$ while hopping in the opposite direction  $-i$.
}\label{fig2}
\end{figure}

We now consider tight binding models for the Majorana zero modes in the triangular and the square vortex lattices. The general Hamiltonian is given in Eq.\ (\ref{kin1}) and the Z$_2$ gauge choice is indicated in Fig.\ \ref{fig2}. Our goal here is do derive the corresponding energy dispersions, assuming nearest neighbor hopping amplitude $t$ for the triangular lattice and both nn and next nn amplitudes $t$ and $t'$ for the square lattice. We will then show in the next Section that such tight binding models accurately describe the MZM dispersions obtained from the full numerical solution of the BdG equations describing the $p_x+ip_y$ superconductor and the Fu-Kane model.

\subsubsection{Square lattice}
If we denote MZMs associated with the two sublattices A and B as $\alpha_\bR$ and $\beta_\bR$ then the Hamiltonian can be written as $\cH_{\square}=\cH_1+\cH_2$ with 
\begin{eqnarray}\label{y1}
\cH_1&=&it \sum_{\bR} \alpha_{\bR}(\beta_{\bR}-\beta_{\bR-\hat{x}-\hat{y}}+\beta_{\bR-\hat{x}}+\beta_{\bR-\hat{y}}), \\
\cH_2&=&it' \sum_{\bR}\left[\alpha_{\bR}(-\alpha_{\bR+\hat{x}}+\alpha_{\bR+\hat{y}})+
\beta_{\bR}(\beta_{\bR+\hat{x}}-\beta_{\bR+\hat{y}})\right]. \nonumber
\end{eqnarray}
This can be diagonalized by passing to the Fourier space
\begin{equation}\label{y2}
\begin{pmatrix}
\alpha_\bR \\
\beta_\bR
\end{pmatrix}
=\sqrt{2\over N} \sum\limits_{\bm k} e^{i(\bm k + \bm Q) \cdot \bm R} \begin{pmatrix}
\alpha_\bk \\
\beta_\bk
\end{pmatrix}, 
\end{equation}
where $\bm Q$ has been inserted for convenience.
We note that $(\alpha^\dagger_\bk,\beta^\dagger_\bk)=(\alpha_{-\bk-2\bm Q},\beta_{-\bk-2\bm Q})$. If we choose $2\bm Q=\bG$ where $\bG$ is a reciprocal lattice vector  
we may restrict $\bk$ to one half of the Brillouin zone and regard $\alpha^\dagger_\bk$ and $\alpha_\bk$ as regular Dirac fermions defined in the reduced BZ. Different choices of $\bm Q$ correspond to different Z$_2$ gauges. The resulting spectra are physically equivalent but may be shifted with respect to the center of the BZ. We adjust $\bm Q$ as necessary to match the spectra obtained in numerical simulations discussed below as we do not apriori know which gauge is chosen by the numerical diagonalization. 

Choosing $\bm Q=(\frac{\pi}{2},-\frac{\pi}{2})$ allows us to use the natural diamond-shaped ``antiferromagnetic'' BZ for this purpose. If we define a two-component spinor $\Gamma_\bk=(\alpha_\bk,\beta_\bk)^T$ the Hamiltonian takes the form $\cH=\sum_\bk\Gamma_\bk^\dagger H_\bk\Gamma_\bk$ with 
\begin{equation}\label{y3}
H_\bk=\begin{pmatrix}
m_\bk & h^*_\bk \\
h_\bk & -m_\bk
\end{pmatrix},
\end{equation}
and 
\begin{eqnarray}\label{y5}
h_{\bm k}&=&-4t e^{i (k_x+k_y)/2}\bigg[\sin \frac{k_x+k_y}{2}-i \sin \frac{k_x-k_y}{2}\bigg], \nonumber \\
m_{\bm k} &=& 4t'(\cos k_x+ \cos k_y).
\end{eqnarray}
The spectrum of excitations
\begin{equation}\label{y4}
E_\bk=\pm\sqrt{|h_\bk|^2+m_\bk^2},
\end{equation}
is gapless with a single Dirac point at $\bk=0$ when $t'=0$ and develops a gap $\Delta=8t'$ otherwise.

\subsubsection{Triangular lattice}
For the triangular vortex lattice in the Z$_2$ gauge indicated in Fig.\ \ref{fig2}(b) the MZM Hamiltonian can be written as $\cH_{\triangle}=\cH_1+\cH_2$ with 
\begin{eqnarray}
\cH_1&=&it \sum_{\bR} \alpha_{\bR}(-\beta_{\bR}+\beta_{\bR-\ba_1-\ba_2}-\beta_{\bR-\ba_1}-\beta_{\bR-\ba_2}), \nonumber \\
\cH_2&=&it \sum_{\bR}\left(-\alpha_{\bR}\alpha_{\bR-\ba_2}+
\beta_{\bR}\beta_{\bR-\ba_2}\right). \label{y6} 
\end{eqnarray}
Here $\ba_1=\sqrt{3}\hat{x}$ and $\ba_2=\hat{y}$ are the primitive
vectors of the sublattice A of the triangular vortex lattice and we
take the distance between the nn\ A vortices as our unit of
length. Fourier transforming according to Eq.\ (\ref{y2}) with the choice ${\bm Q}=(0,0)$ leads to the Bloch Hamiltonian of the form indicated in Eq.\ (\ref{y3}) with
\begin{eqnarray}\label{y7}
h_{\bm k}&=&4t e^{{i\over 2} (\sqrt{3}k_x+k_y)}\bigg[\sin \frac{\sqrt{3}k_x+k_y}{2}+i \cos \frac{\sqrt{3}k_x-k_y}{2}\bigg], \nonumber \\
m_{\bm k} &=& 4t \sin{k_y}.
\end{eqnarray}
The spectrum has the form of Eq.\ (\ref{y4}) and is fully gapped for the triangular lattice. The smallest excitation energy $4t$ attains at the $\Gamma$ point while the maximum $4\sqrt{3}t$ occurs along the line between $\Gamma$ and $M$ points of the Brillouin zone.


\section{Numerical solutions of the BdG problem with vortices}

Our objective in this Section is to find numerical solutions of the full BdG equation (\ref{z2}) in the presence of a vortex lattice (and the accompanying magnetic field $\bB$). We do this for both the continuum and the lattice formulations of a model $p_x+ip_y$ superconductor as well as the Fu-Kane model. We use these numerical solutions to ascertain the validity of the low-energy effective theories for the Majorana fermions derived in Sec.\ II and to relate the parameters that enter these theories to the physical parameters characterizing the microscopic models.

\subsection{Continuum formulation}
Here we wish to solve the BdG equation (\ref{z2}) for Hamiltonians (\ref{z1}) and (\ref{fk1}) defined in the continuum. The vortex lattice is encoded in the SC phase field $\theta(\br)$ as described in Eq.\ (\ref{z3}) and the magnetic field is included through the minimal substitution. The key difficulty in solving the BdG equation under these conditions lies in the fact that although we expect the physical observables to exhibit periodicity of the underlying vortex lattice the Hamiltonian itself is not periodic. As noted originally in Refs.\ \cite{tesanovic1,vafek1} this difficulty can be circumvented by performing a singular gauge transformation  
\begin{equation}\label{n1}
H\to \tilde{H}=UHU^{-1}, \ \ U=
\begin{pmatrix}
e^{-i\theta_A(\br)} & 0 \\
0 & e^{i\theta_B(\br)}
\end{pmatrix}
\end{equation}
where $\theta_A(\br)$ and $\theta_B(\br)$ are two functions satisfying
\begin{equation}\label{n2}
\theta_A(\br)+\theta_B(\br) = \theta(\br),
\end{equation}
and are chosen such that $U(\br)$ defined above is single-valued. In practice this is achieved by partitioning vortices into two sublattices $A$ and $B$ and assigning the contribution from sublattice $A$ to $\theta_A(\br)$ and sublattice $B$ to $\theta_B(\br)$. The transformed Hamiltonian $\tilde{H}$ is then periodic and single valued \cite{tesanovic1,vafek1} and can be analyzed using the standard band structure techniques.

\subsubsection{$p_x+ip_y$ superconductor}

For the $p_x+ip_y$ SC the transformed Hamiltonian reads
\begin{equation}\label{n3}
\tilde{H}=
\begin{pmatrix}
{1\over 2m}(\bp+\bv_s^A)^2-\mu & \tilde\Delta_+ \\
\tilde{\Delta}_- & -{1\over 2m}(\bp-\bv_s^B)^2+\mu
\end{pmatrix},
\end{equation}
where $\tilde\Delta_\pm=\tilde\Delta_x\pm i\tilde\Delta_y$ and 
\begin{equation}\label{n4}
\tilde{\bm\Delta}=\Delta_0\left[\bp+{1\over 2}(\bv_s^A-\bv_s^B)\right].
\end{equation}
Furthermore, quantities 
\begin{equation}\label{n5}
\bv_s^\mu=\nabla\theta_\mu-{e\over c}\bA, \ \ \ \mu=A,B
\end{equation}
are related to the physical superfluid velocity
$\bv_s=\bv_s^A+\bv_s^B$ and are gauge invariant as well as periodic in
real space. One can, at  least in principle, solve the eigenvalue
problem defined by $\tilde{H}$ by exploiting the Bloch theorem  and going to the momentum space. Here, following Read and Green \cite{read1}, we consider a slightly simpler problem that follows from sending $m\to\infty$ in Hamiltonian (\ref{n4}). As argued in Ref.\ \cite{read1} one expects this limit to show the same qualitative behavior as the full model: the system is in the topological phase with unpaired MZMs in vortex cores in the ``weak pairing'' phase that obtains when $\mu>0$ and is in the trivial ``strong pairing'' phase otherwise.  

With the above considerations in mind we model vortex core as a small circular region tuned to the trivial phase by locally setting $\mu$ large and negative. The MZM can then be pictured as a chiral edge state at the boundary between the topological bulk and the trivial core region.
 This treatment of the vortex core also circumvents a difficulty that is known to arise in numerical solutions of problems with Dirac Hamiltonians in continuum. As first noted in Ref.\ \cite{vafek1} and later elaborated in Ref.\ \cite{ashot1} eigenstates of the Dirac Hamiltonian tend to diverge as $\sim 1/\sqrt{r}$ in the vicinity of the vortex core resulting an an overcomplete basis of states. (Note that this behavior is characteristic of the $m\to\infty$ approximation.) To treat this problem one must regularize the theory in some fashion at short distances. Modeling the core as a trivial strong pairing region represents one possible way to regularize by suppressing the wavefunctions at the core center. In the next subsection we will discuss the lattice formulation of the model which provides another natural regularization scheme. Importantly, we shall see that the low energy properties of the system (i.e.\ the effective  theory for the MZMs) are independent of the details of the regularization scheme.

The problem we solve numerically is therefore defined by the Hamiltonian
\begin{equation}\label{n6}
\tilde{H}(\br)=
\begin{pmatrix}
-\mu(\br) & \tilde\Delta_+ \\
\tilde{\Delta}_- & \mu(\br)
\end{pmatrix},
\end{equation}
where $\tilde{\bm\Delta}$ is given in Eq.\ (\ref{n4}).  The superfluid velocities $\bv_s^\mu$ that enter the gap function can be determined by a procedure analogous to that leading to Eq.\ (\ref{x7}) above (see also Appendix B in Ref.\ \cite{vafek1} for a more detailed description).
The chemical potential is taken as 
\begin{equation}\label{n7}
\mu(\br)=\mu_0-\mu'\sum_je^{-(\br-\bR_j)^2/\xi^2}
\end{equation}
with the second term representing the vortex cores, as discussed
above. The specific Gaussian form is not important (any functional
form peaked at $\br=\bR_j$ would work) but is convenient for the
numerics because it has a simple Fourier transform. With these
preparations we can now employ the Bloch theorem, Fourier transform
the Hamiltonian (\ref{n6}) as described e.g.\ in Ref.\
\cite{tesanovic1}, and find its energy eigenvalues for each crystal
momentum $\bk$ in the first Brillouin zone by a straightforward
numerical diagonalization. Because the spectrum of the continuum
Hamiltonian (\ref{n7}) is unbounded we have to impose a high energy
cutoff $\Lambda$ to render the Bloch matrix finite. $\Lambda$ must be
chosen sufficiently large so that the low-energy spectrum no longer
depends on it.

\begin{figure}[t]
\includegraphics[width = 8.0cm]{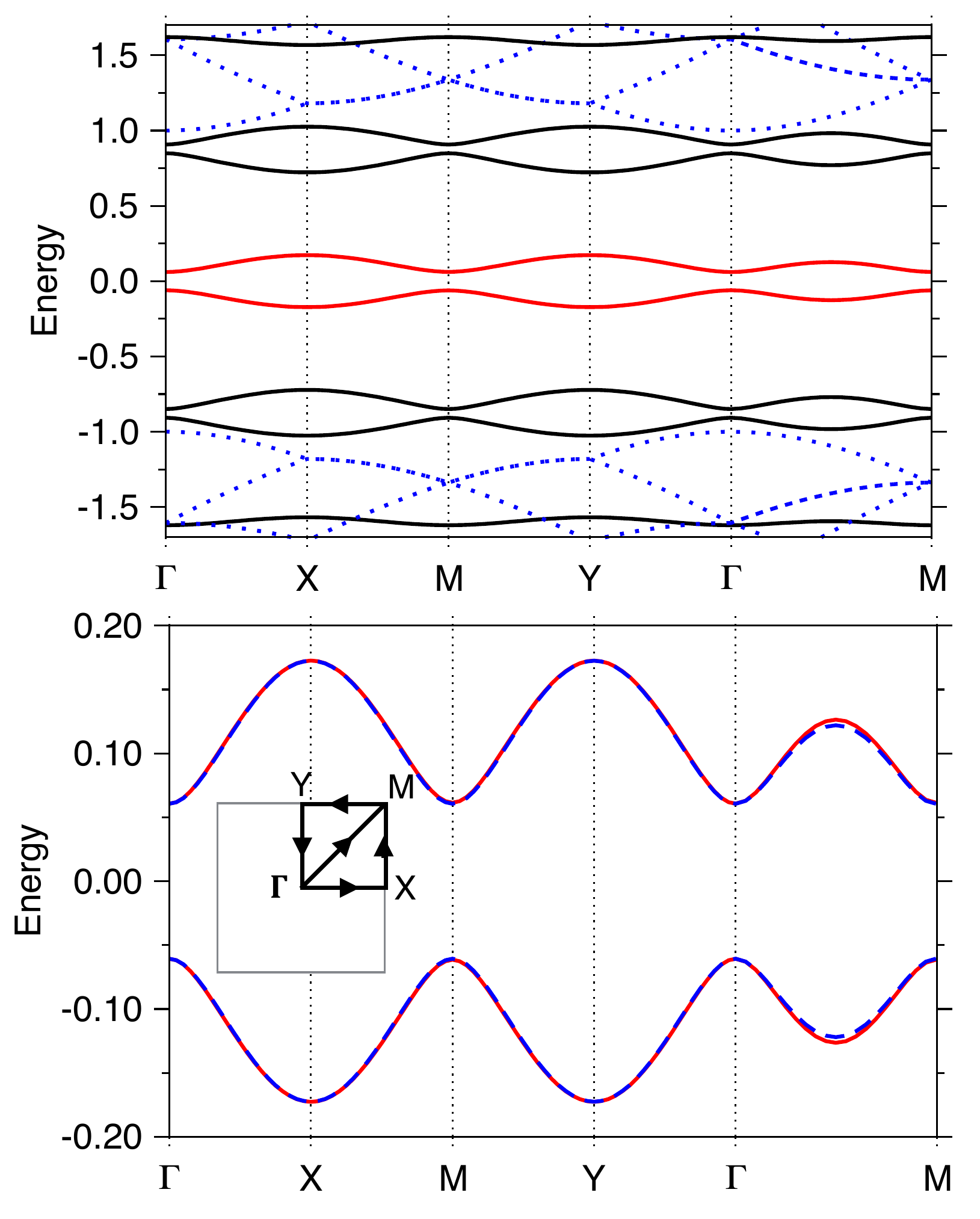}
\caption{Band structure for the continuum model of a $p_x+ip_y$ superconductor. In the top panel solid (dotted) lines indicate the band structure calculated in the presence (absence) of the square vortex lattice. The parameters are chosen as follows, $\mu_0=\Delta_0=1$, $\mu'=40$ and $\xi/a=0.2$. The two bands closest to zero energy are the MZM bands. They are enlarged in the bottom panel where the dashed line represents the best fit to the Majorana tight binding model   
(\ref{y4}) with the hopping parameters $t=0.0305$ and $t'=0.0076$. The inset shows the path taken in the first Brillouin zone.
}\label{fig3}
\end{figure}
Typical results for the square vortex lattice are displayed in Fig.\
\ref{fig3}. In the absence of vortices the spectrum shows a
gap $\mu_0$. When vortices are present states appear inside the gap. These are the expected vortex core bound states broadened into bands by intervortex hybridization. The
pair of bands closest to zero energy  are formed of MZMs. We checked that these bands become flat and
approach zero energy in the limit of a dilute vortex lattice $a\gg
\xi$. Their dispersion shows an excellent agreement with the tight
binding model for MZMs Eq.\ (\ref{kin1}) with the Z$_2$ gauge factors
given by the  Grosfeld-Stern rule Eq.\ (\ref{kin2}).

\subsubsection{Fu-Kane model}

After the singular gauge transformation (\ref{n1}) the Fu-Kane Hamiltonian (\ref{fk1}) takes the form 
\begin{equation}\label{n8}
\tilde{H}_{\rm FK}(\br)=
\begin{pmatrix}
v\sigma\cdot(\bp+\bv_s^A)-\mu & \Delta_0 \\
\Delta_0 & -v\sigma\cdot(\bp-\bv_s^B)+\mu 
\end{pmatrix},
\end{equation}
where the $\bv_s^\mu$ velocities are given by Eq.\ (\ref{n5}) as before. Once again, the transformed Hamiltonian is periodic and single valued. Unlike the Hamiltonian for the $p_x+ip_y$ SC which has two distinct phases depending on the sign of $\mu$ the Fu-Kane Hamiltonian remains in the same (topological) phase for all values of $\mu$ when $\Delta_0$ is non-zero. In order to regularize the wavefunction behavior in the vortex cores we thus introduce a small modification $\tilde{H}_{\rm FK}\to \tilde{H}_{\rm FK}+\delta H_m$ to the Hamiltonian (\ref{n8}), making the core magnetic using    
\begin{equation}\label{n9}
\delta H_m(\br)=
\begin{pmatrix}
\sigma^z m_z(\br) & 0\\
0 &  \sigma^z m_z(\br)
\end{pmatrix}.
\end{equation}
As before we take $m(\br)$ to be large in the vortex cores and zero outside; specifically  
\begin{equation}\label{n10}
m(\br)=m_0\sum_je^{-(\br-\bR_j)^2/\xi^2},
\end{equation}
and identify $\xi=v/\pi\Delta_0$ with the SC coherence length.
Magnetic order breaks the time reversal symmetry of the TI surface state and is known to gap out the protected gapless states. The MZMs then can be viewed as edge states living on the boundary between the predominantly magnetic core region and the SC bulk. It is, however,  important to emphasize that without the long-ranged phase structure due to vortices encoded in the $\bv_s^\mu$ factors the perturbation (\ref{n9}) by itself would {\em not} produce MZMs as the edge modes would exhibit a large finite size gap $\sim v/\xi\sim\Delta_0$. It is the phase structure that is instrumental for the emergence of MZMs whereas $m_z(\br)$ serves merely to regularize the continuum theory at short distances.
In the next subsection we will see that the same MZM structures arise from a theory regularized on the lattice where there is no need to include the magnetic order.
\begin{figure}[t]
\includegraphics[width = 8.0cm]{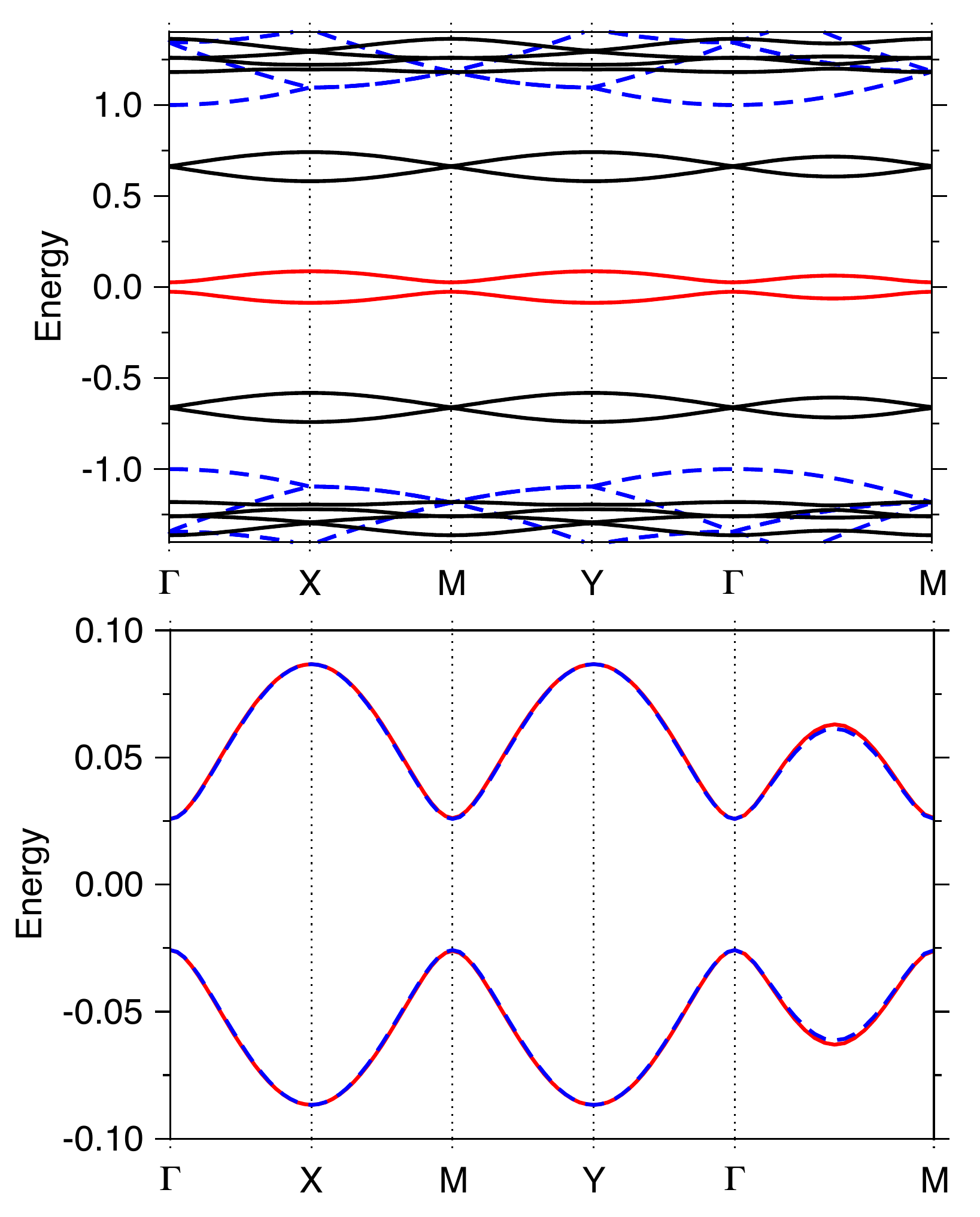}
\caption{Band structure for the continuum Fu-Kane model, square vortex lattice. In the top panel solid (dashed) lines indicate the band structure calculated in the presence (absence) of the vortex lattice. The parameters are chosen as follows: $v=0.14$, $m_0=7.0$, $\mu=0$ and $\xi/a=0.14$. The high energy cutoff $\Lambda=8$ and all the quantities are in units of $\Delta_0=1$. As in Fig.\ \ref{fig4} the two bands closest to zero energy are the MZM bands. They are enlarged in the bottom panel where the dashed line represents the best fit to the Majorana tight binding model   
(\ref{y4}) with the hopping parameters $t=0.0153$ and $t'=0.0032$. 
}\label{fig4}
\end{figure}

\begin{figure}[t]
\includegraphics[width = 8.0cm]{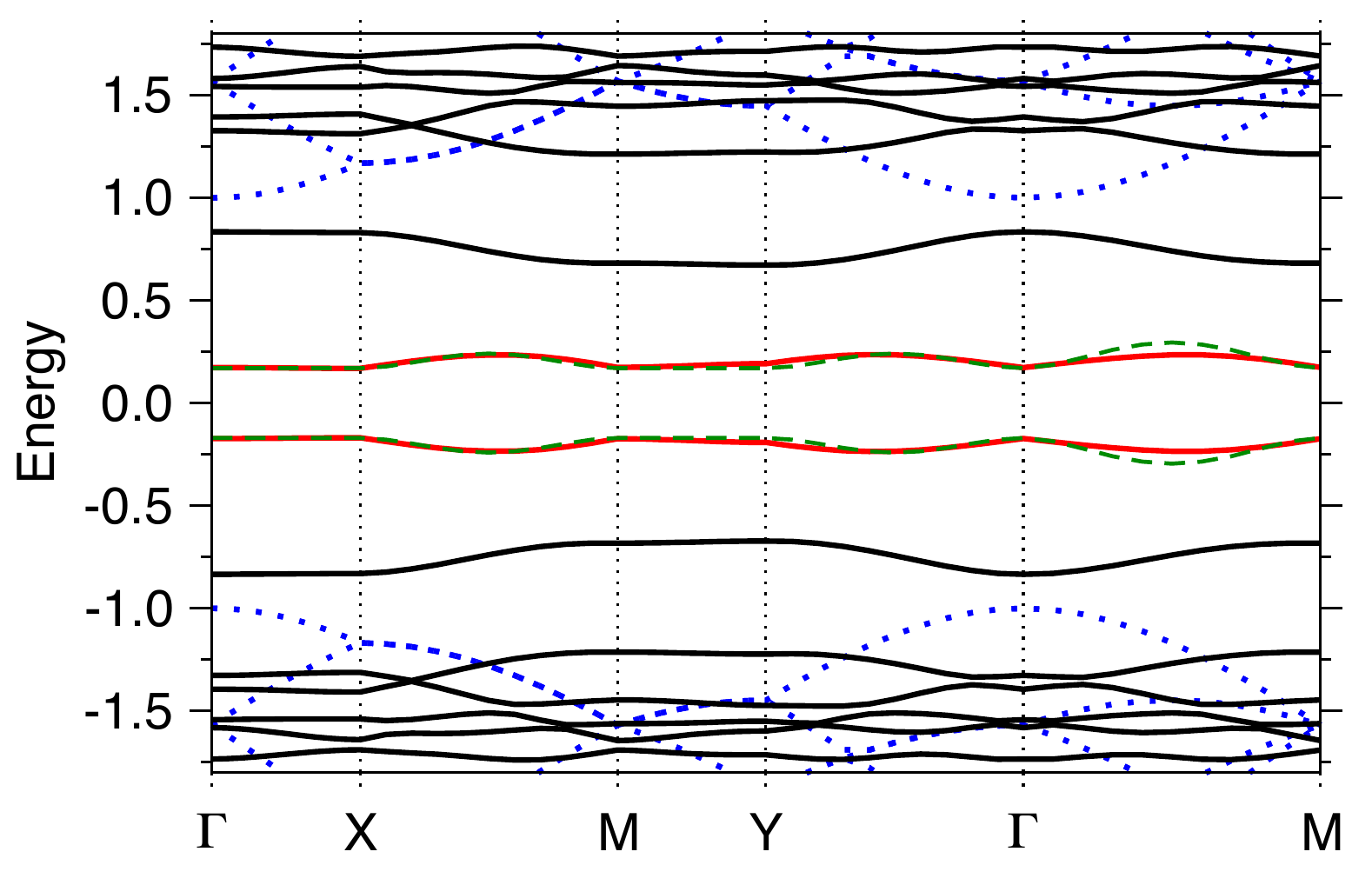}
\caption{Band structure for the continuum Fu-Kane model, triangular vortex lattice. Solid (dotted) lines indicate the band structure calculated in the presence (absence) of the vortex lattice. The parameters are chosen as follows: $v=0.33$, $m=1.0$, $\mu=0$ and $\xi/a=0.17$. The high energy cutoff $\Lambda=20$ and all the quantities are in units of $\Delta_0=1$. As in Fig.\ \ref{fig4} the two bands closest to zero energy are the MZM bands. The dashed line represents the best fit to the Majorana tight binding model   
(\ref{y6}) with the hopping parameter $t=0.0425$. 
}\label{fig5}
\end{figure}

With this preparation it is now straightforward to numerically diagonalize the Bloch Hamiltonian $\tilde{H}_{FK}(\bk)$ that follows from Eqs.\ (\ref{n8}-\ref{n10}) upon Fourier transforming and imposing the high energy cutoff $\Lambda$ to render the Bloch matrix finite. Typical results of such a calculation are displayed in Figs.\ \ref{fig4} and \ref{fig5} for the square and the triangular vortex lattices, respectively. In both cases we observe the initially gapped spectrum (in the absence of vortices) modified by the emergence of the low energy vortex core states. The bands closest to zero energy arise from MZMs. For the square vortex lattice their dispersions show near perfect agreement with the Majorana tight binding models derived in Sec.\ II. For the triangular vortex lattice the agreement is also good and can be further improved by including longer range hoppings. Since the simplest nn tight binding model already captures all the qualitative features of the MZM band we do not pursue this here.

We conclude this subsection by noting the qualitative and quantitative similarity between the MZM bands found in the $p_x+ip_y$ SC and the Fu-Kane model. Indeed this is not surprising in view of the expectation that they be described by the same minimal tight binding model with static Z$_2$ gauge structure described in Sec.\ II. The one distinguishing feature of the Fu-Kane mode -- the flat MZM  bands  expected at $\mu=0$ due to the extra chiral symmetry -- is not apparent in the continuum formulation. This is because the $\delta H_m$ term introduced to regularize the continuum model breaks the chiral symmetry (even at $\mu=0$). We were unable to find a symmetry preserving regulator that would work for this purpose in the continuum model. We shall see however that the lattice model considered next preserves the chiral symmetry and indeed exhibits the expected flat bands at zero energy.

\subsection{Lattice formulation}
Although technically somewhat more complicated the lattice formulation of the problem has a distinct advantage of providing a natural short distance cutoff for the electron wavefunctions in vortex cores. Artificial regulators that were necessary in the continuum theory are thus not needed. The lattice formulation of the $p_x+ip_y$ superconductor with vortices has been discussed in Ref.\ \cite{vafek1}  although Majorana bands have not been studied in detail. Here we briefly review the lattice construction and examine the Majorana bands. This we follow by a similar discussion for the Fu-Kane model whose lattice formulation has not been previously discussed.
\begin{figure}[t]
\includegraphics[width = 8.0cm]{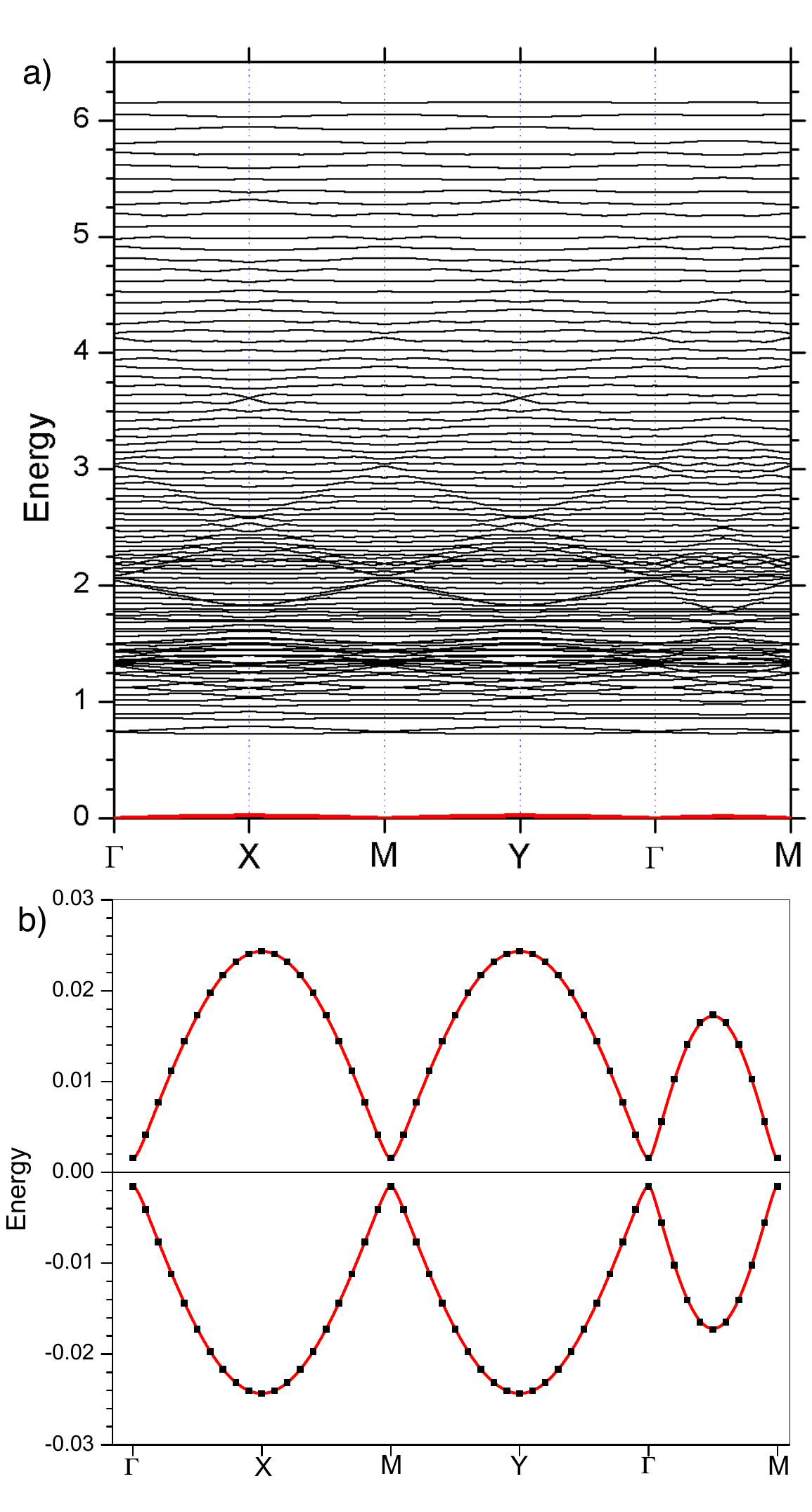}
\caption{Band structure for the lattice $p_x+ip_y$ superconductor with a square vortex lattice. a) Full band structure in a $10\times 10$ magnetic unit cell with $\Delta_0=0.50$ and $\varepsilon_F=-2.2$. b) Detail of the Majorana band (solid squares) and the best fit to the tight-binding Majorana dispersion Eq.\ (\ref{y4}) with $t=4.3027\times 10^{-3}$ and $t'=1.9375\times 10^{-4}$ (solid line).   
}\label{fig6}
\end{figure}
%

\subsubsection{$p_x+ip_y$ superconductor}
The problem is defined by the Hamiltonian (\ref{z0},\ref{z1}) with the fermion creation operators now residing on sites of a lattice which we take to be square and of unit spacing. The kinetic and pairing operators are given by  \cite{vafek1}
\begin{eqnarray} \label{w1}
	\hat{h}&=&-\tau \sum_{\bm{\delta}}e^{-i(e/ \hbar c) \int^{\bm{r}+\bm{\delta}}_{\bm{r}}\bm{A}(\bm{r})\cdot d \bm{l}} 		  \hat{s}_{\bm{\delta}}-\varepsilon_F,
\\ \label{w2}
	\hat{\Delta}&=&\Delta_0 \sum_{\bm{\delta}}e^{i \theta(\bm{r})/2}\hat{\eta}_{\bm{\delta}}e^{i \theta(\bm{r})/2}.
\end{eqnarray}
From now on, we will set the hopping amplitude \(\tau \)  to unity and measure all energies in units of \(\tau \). Also, \( \hat{s}_{\bm{\delta}}\) denotes the shift operator \( \hat{s}_{\bm{\delta}}u(\bm{r})=u(\bm{r}+\bm{\delta})\), where ${\bm \delta}$ represents a nn vector. For the $p_x+ip_y$ superconductor the operator \( \hat{\eta}_{\bm{\delta}} \) is defined as
\begin{equation}\label{w3}
	\hat{\eta}_{\bm{\delta}}=
	\begin{cases}
	\mp i \hat{s}_{\bm{\delta}} &\mbox{ if \(\bm{\delta}=\pm \hat{x}\)},\\
	\pm \hat{s}_{\bm{\delta}} &\mbox{ if \(\bm{\delta}=\pm \hat{y}\)}.
	\end{cases}
\end{equation}
After the singular gauge transformation (\ref{n1}) we obtain the Hamiltonian 
\begin{equation}\label{w4}
\tilde{H}=\begin{pmatrix}
		 - \sum\limits_{\bm{\delta}}e^{i \mathcal{V}_{\bm{\delta}}^A (\bm{r})}\hat{s}_{\bm{\delta}}-\varepsilon_F&
		 \Delta_0 \sum\limits_{\bm{\delta}}e^{i \mathcal{A}_{\bm{\delta}} (\bm{r})}  \hat{\eta}_{\bm{\delta}}	\\
		\Delta_0 \sum\limits_{\bm{\delta}}e^{i \mathcal{A}_{\bm{\delta}} (\bm{r})}  \hat{\eta}_{\bm{\delta}}^*& 
		 \sum\limits_{\bm{\delta}}e^{-i \mathcal{V}_{\bm{\delta}}^B (\bm{r})}\hat{s}_{\bm{\delta}}+\varepsilon_F
	\end{pmatrix}
\end{equation}
with the phase factors defined as
\begin{equation}\label{w5}	\mathcal{V}_{\bm{\delta}}^{\mu}(\bm{r})=\int_{\bm{r}}^{\bm{r}+\bm{\delta}}\Big(\nabla \theta_{\mu}-\frac{e}{\hbar c} \bm{A}\Big) \cdot d \bm l \qquad \mu=A,B
\end{equation}
and $\mathcal{A}_{\bm{\delta}}(\bm{r})= \frac{1}{2}[\mathcal{V}_{\bm{\delta}}^A (\bm{r}) - \mathcal{V}_{\bm{\delta}}^B (\bm{r})]$. The phase factors are easily evaluated using the method discussed in Sec.\ II.D (see also Appendix B in Ref.\ \cite{vafek1} for details). 

 The Hamiltonian (\ref{w4}) now has the periodicity of the vortex lattice (with two vortices per unit cell) and can be diagonalized in momentum space using standard band structure techniques. Fig.\ \ref{fig6}(a) shows the band structure obtained for the square vortex lattice. It exhibits the expected Majorana band close to zero energy as well as nearly flat Landau level bands at energies high compared to the SC gap amplitude $\Delta_0$, in complete agreement with results of Ref.\ \cite{vafek1}. Panel (b) of the figure focuses on the Majorana band which is, once again, very well described by the tight-binding dispersion (\ref{y4}).
\begin{figure}[t]
\includegraphics[width = 8.0cm]{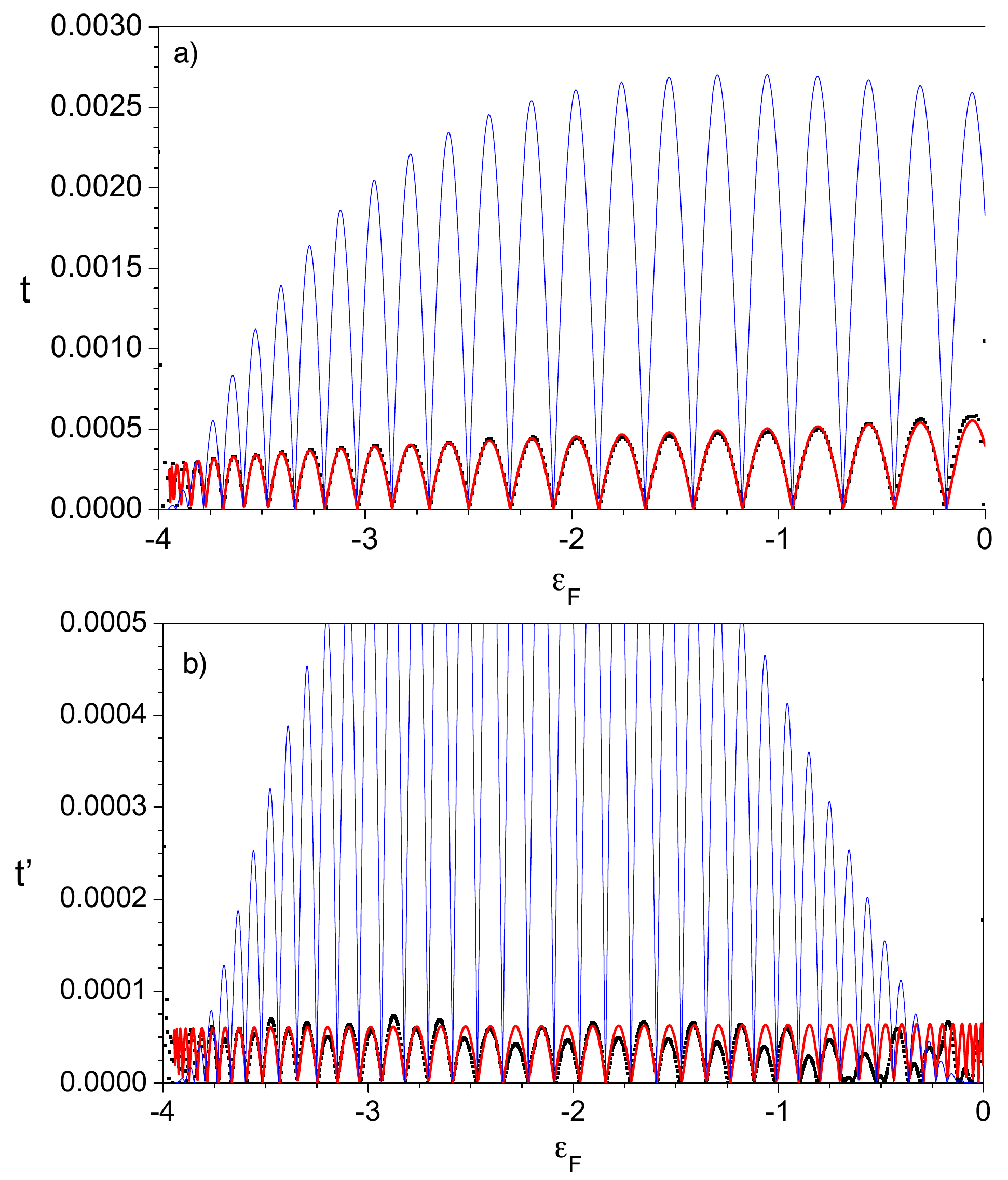}
\caption{The Majorana overlap amplitudes $t$ and $t'$ as a function of $\varepsilon_F$ extracted from the lattice model $p_x+ip_y$ superconductor (solid symbols). The magnetic unit cell is $50\times 50$ and $\Delta_0=0.50$. The thin (blue) line represents the analytical result of Ref.\ \cite{galitsky2} while the thick (red) line corresponds to the simple phenomenological expression (\ref{w7}) discussed in the text. The fit parameters are $A=2.43\times 10^{-4}$, $b=0.0163$ and $A'=5.91\times 10^{-5}$, $b'=0.0066$.
}\label{fig7}
\end{figure}

From these results we may easily extract the dependence of the tunneling amplitudes $t$ and $t'$ that enter the effective tight-binding Majorana model on various microscopic parameters of the underlying BdG theory as well as the vortex lattice geometry. We do this by fitting the numerically calculated MZM bands to the tight binding dispersion (\ref{y4}).  As an example Fig.\ \ref{fig7} displays the dependence of $(t,t')$ on the Fermi energy $\varepsilon_F$ with $\Delta_0$, $\tau$ and vortex spacing held fixed. 
We note that in view of Eq.\ (\ref{y4}) the Majorana band structure is not sensitive to the sign of the amplitudes $t$ and $t'$. We may plausibly surmise that the nodes apparent in  Fig.\ \ref{fig7} represent sign changes in the amplitudes which fixes them up to an overall sign. The overall sign could be potentially determined from the structure of the corresponding wavefunctions but we do not pursue this issue here since we do not believe the sign is an easily measurable quantity.

We can compare these with the analytical expressions derived in Refs.\ \cite{galitsky1,galitsky2}. Specifically, we plot 
\begin{equation}\label{w6}
t\approx \sqrt{\frac{2}{\pi}} \Delta_0 \frac{|\cos{(k_F R + {\pi\over 4})}|}{\sqrt{k_F R}}\exp\Big(-\frac{R}{\xi}\Big)
\end{equation}
which corresponds to Eq.\ (31) of Ref.\ \cite{galitsky2}, valid in the weak coupling limit $\Delta_0\ll\varepsilon_F$ with $\varepsilon_F$ referenced to the bottom of the band. Here $R$ denotes the distance between the vortices, $k_F$ is the Fermi momentum and $\xi=v_F/\pi\Delta_0$ is the BCS coherence length. The above expression accurately captures the period and the phase of oscillations in both $t$ and $t'$ but does not describe the amplitude particularly well. We tried other, more complicated expressions derived in Ref.\ \cite{galitsky2}, but they do not significantly improve the agreement. We instead find that the data is well described by a simple phenomenological expression
\begin{equation}\label{w7}
t\approx A(1+bRk){|\cos{(k R + {\pi/4})}|}.
\end{equation}
Here $k=\sqrt{k_F^2-(\Delta_0/v_F)^2}$ while $A$ and $b$ are dimensionless constants, and a similar expression for $t'$ with parameters $A'$ and $b'$. 
Because the MZM wavefunctions decay exponentially with the characteristic lengthscale $\xi$ the amplitude implied by the expression (\ref{w6}) makes good intuitive sense. Our results suggest that the interplay between MZM wavefunctions in the vortex lattice is possibly quite intricate and cannot be fully captured by the perturbative treatment of two distant vortices carried out in Refs.\ \cite{galitsky1,galitsky2}.

\subsubsection{Fu-Kane model}
The implementation of the Fu-Kane model on the lattice is more involved owing to the Nielsen-Ninomyia theorem \cite{NN}, which states that it is
impossible, as a matter of principle, to construct a $\cal{T}$-invariant
2D lattice Hamiltonian with an odd number of Dirac fermions in the
low-energy spectrum. It is therefore impossible to write a 2D lattice model that would faithfully describe a single surface of a TI. 
Studying the full 3D problem (including the STI bulk)
  would provide the desired outcome but this would be computationally
  very costly. We also note that magnetic field of several Tesla,
  sufficient to produce the vortex lattice, has negligible effect on
  the gapped bulk of the STI. This is because the relevant cyclotron
  frequency as well as the Zeemann energy are much smaller than the
  bandgap (which is $\sim 300$ meV in Bi$_2$Se$_3$ family of
  materials). There is, therefore, nothing interesting to learn by
  performing a full 3D calculation. The problem of doped STI, where the
bulk itself can become superconducting, has been studied with some
interesting results \cite{hughes1,hughes2}.

 To circumvent the above problem we employ the idea introduced in Ref.\ \cite{marchand1} and construct a lattice model describing instead a pair of parallel TI surfaces, such as those terminating a slab. Because a pair of TI surfaces has in general an even number of Dirac fermions the theorem \cite{NN} no longer presents an obstruction.  
Ref.\ \cite{marchand1} showed how to construct a lattice model of this type with low-energy degrees of freedom on two surfaces that are largely decoupled.

The normal state Hamiltonian (``model II'' in Ref.\ \cite{marchand1}) can be written in the momentum space as 
\begin{equation}\label{w10}
	h_\bk= \begin{pmatrix}
		g_{\bm k} & \bar{M}_{\bm k}	\\
		\bar{M}_{\bm k}&-g_{\bm k}
	\end{pmatrix}
\end{equation}
with $g_{\bm k}=2 \lambda (\sigma^y \sin k_x-\sigma^x \sin k_y)$ and  
$\bar{M}_{\bm k}=2\tau (2- \cos k_x - \cos k_y)$. Its diagonal blocks describe the gapless surface states in the two surfaces of a TI and the coupling $ \bar{M}_{\bm k}$ is designed to gap out all the Dirac nodes except those at the origin, $\bk=(0,0)$. In the following we set $\lambda=1$ and measure all energies in units of $\lambda$.
To study the vortex lattice we imagine inducing superconductivity in one of the surfaces by proximity effect. This is described by passing to the BdG formulation using Eq.\ (\ref{fk1}) with 
\begin{equation}\label{w11}
\hat{\Delta}=\begin{pmatrix}
\Delta & 0 \\
0 & 0
\end{pmatrix}.
\end{equation}
To avoid complications that would arise from the other surface being ungapped we imagine that its surface state has been gapped by a $\cal{T}$-breaking perturbation and replace $g_\bk\to g_\bk+m\sigma^z$ in the lower diagonal element of Eq.\ (\ref{w10}). Since the physical time-reversal is already broken by the applied magnetic field we do not expect this additional $\cal{T}$-breaking to have a significant effect on the system, other than removing the unwanted gapless excitations from the second surface. We also emphasize that this is a purely technical device and we do not require such magnetization to be implemented in the experimental realization.
\begin{figure}[t]
\includegraphics[width = 8.0cm]{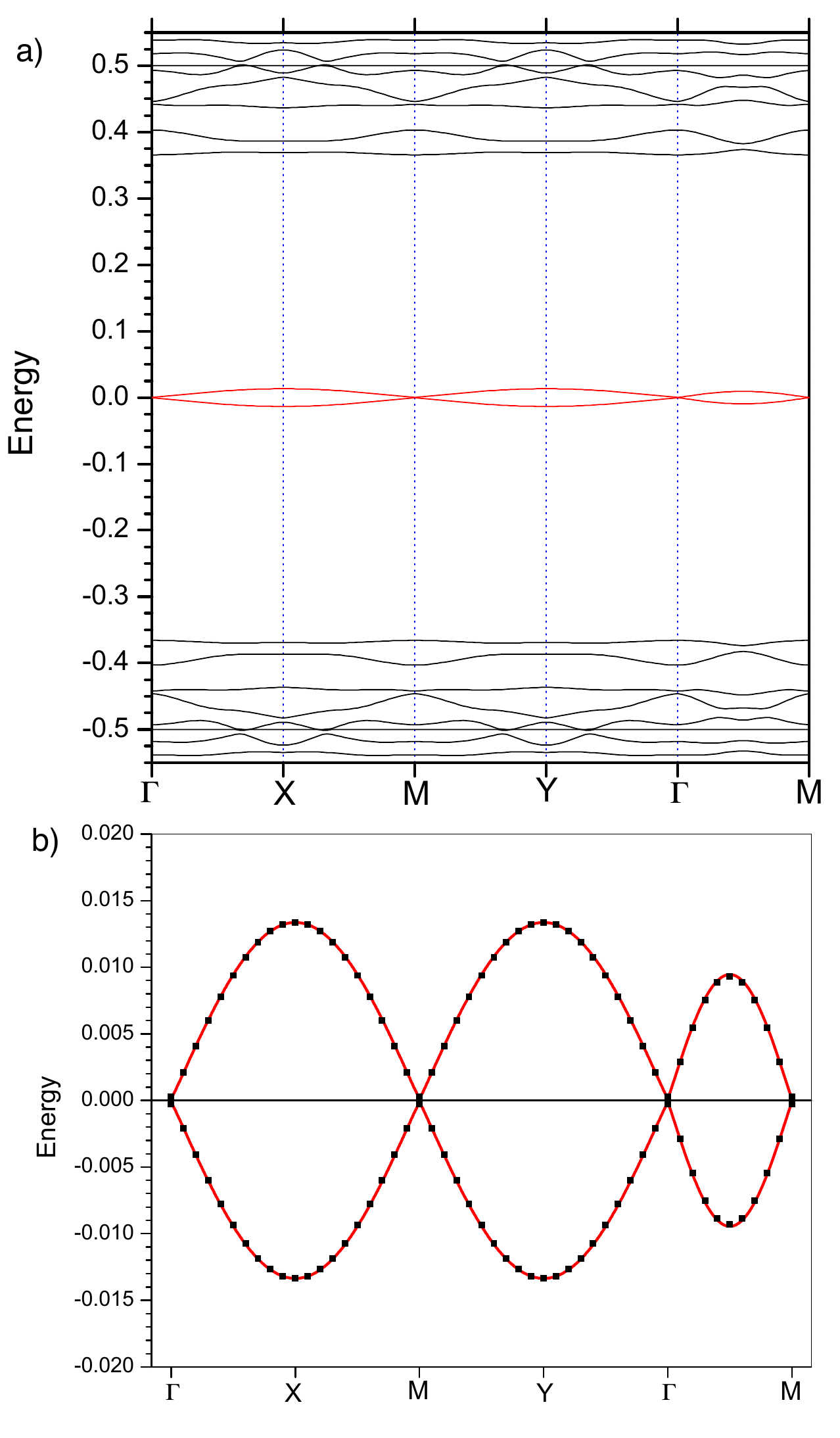}
\caption{Band structure for the lattice version of the Fu-Kane model with a square vortex lattice. a) Full band structure in a $30\times 30$ magnetic unit cell with $\tau=0.5$,  $\Delta_0=0.4$, $m=0.5$ and $\varepsilon_F=0.25$. b) Detail of the Majorana band (solid squares) and the best fit to the tight-binding Majorana dispersion Eq.\ (\ref{y4}) with $t=2.36\times 10^{-3}$ and $t'=3.45\times 10^{-4}$ (solid line).   
}\label{fig8}
\end{figure}

The full momentum-space Hamiltonian we consider thus has the following form
\begin{equation}\label{w12}
H^{FK}_\bk= \begin{pmatrix}
g_{\bm k}-\varepsilon_F   & \bar{M}_{\bm k} & \Delta & 0	\\
\bar{M}_{\bm k} & -g_{\bm k}-m\sigma^z & 0 & 0    \\
\Delta^*       & 0 & -g_{\bm k} + \varepsilon_F & -\bar{M}_{\bm k} \\
0 & 0 &      -\bar{M}_{\bm k} & g_{\bm k}-m\sigma^z
	\end{pmatrix}.
\end{equation}
In order to implement the vortex lattice we now pass to the real space and perform the minimal substitution to include the magnetic field. The upper diagonal block of $H^{FK}$ thus becomes
\begin{equation}\nonumber
\begin{pmatrix}
	
		-\varepsilon_F & i \sum\limits_{\bm \delta} \hat{\eta}_{\bm \delta}^*&
		4\bar{t}-\bar{t} \sum\limits_{\bm \delta} \hat{s}_{\bm \delta}& 0 \\
		i \sum\limits_{\bm \delta} \hat{\eta}_{\bm \delta} & -\varepsilon_F & 0
		& 4\bar{t}-\bar{t} \sum\limits_{\bm \delta} \hat{s}_{\bm \delta}\\
		4\bar{t}-\bar{t} \sum\limits_{\bm \delta} \hat{s}_{\bm \delta} & 0 & -m &
		-i \sum\limits_{\bm \delta} \hat{\eta}_{\bm \delta}^*\\
		0 & 4\bar{t}-\bar{t} \sum\limits_{\bm \delta} \hat{s}_{\bm \delta} &
		-i \sum\limits_{\bm \delta} \hat{\eta}_{\bm \delta} & m	
	
	\end{pmatrix}
\end{equation} 
and a similar expression for the lower diagonal block. Vortices are included by replacing $\Delta\to\Delta e^{i\theta(\br)}$ and the magnetic field enters via the Peierls substitution 
\begin{equation}\label{w13}
\hat{s}_{\bm\delta}\to e^{-i(e/ \hbar c) \int^{\bm{r}+\bm{\delta}}_{\bm{r}}\bm{A}(\bm{r})\cdot d \bm{l}} 		  \hat{s}_{\bm{\delta}}.
\end{equation} 
We note that, importantly, the Peierls phase factors must now be also attached to the shift operators that enter the definition of $\hat{\eta}_{\bm \delta}$ because in the Fu-Kane model these appear in the kinetic energy of the system and thus represent single-electron hopping processes.   
\begin{figure}[t]
\includegraphics[width = 8.0cm]{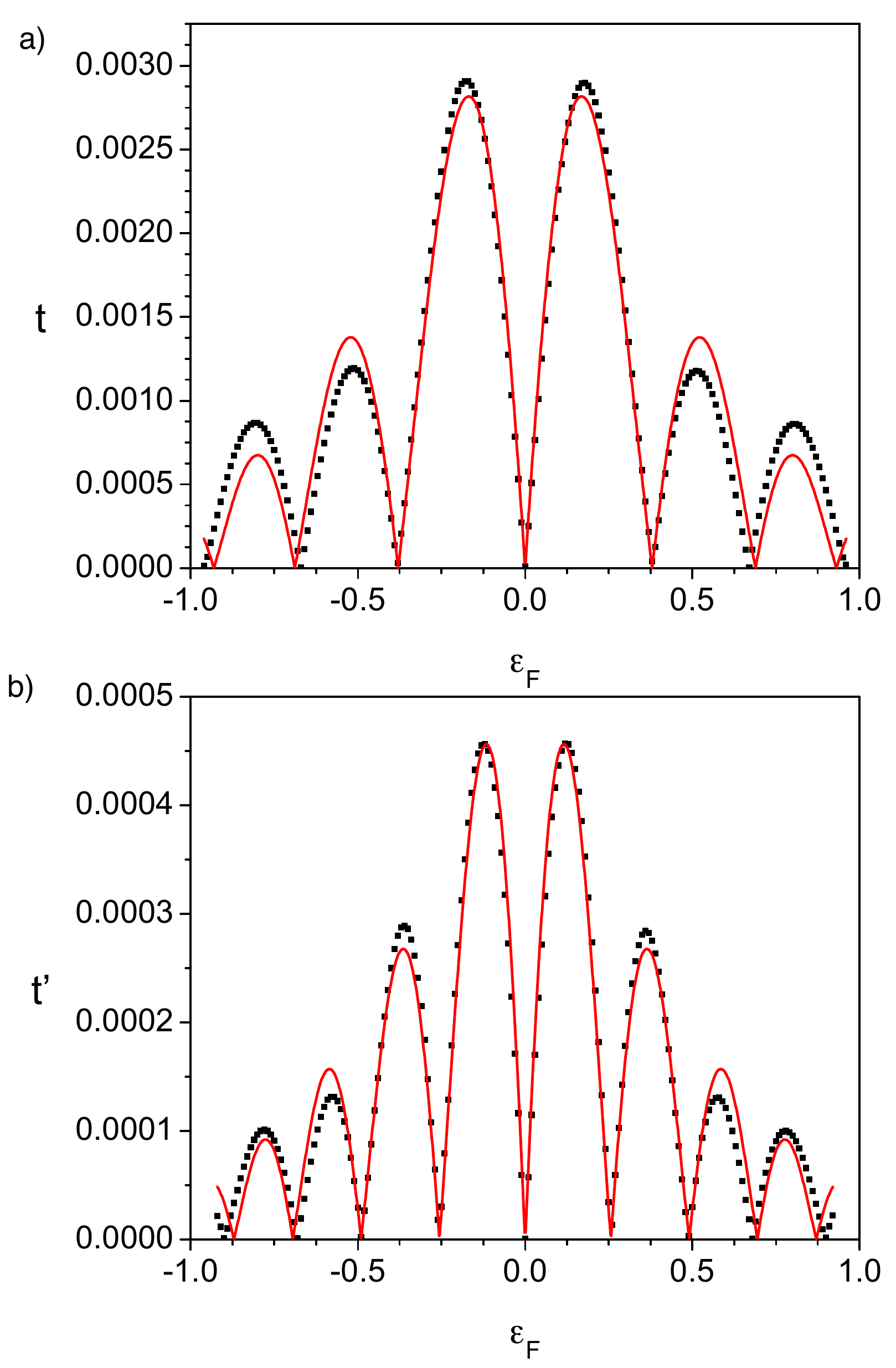}
\caption{The Majorana overlap amplitudes $t$ and $t'$ as a function of $\varepsilon_F$ extracted from the Fu-Kane model formulated on the lattice (solid symbols). The magnetic unit cell is $30\times 30$, $\tau=0.5$, $\Delta_0=0.4$ and $m=0.5$. The solid (red) line corresponds to the simple phenomenological expression (\ref{w13}) discussed in the text. The fit parameters are $A=7.85\times 10^{-3}$, $b=0.224$ and $A'=5.87\times 10^{-4}$, $b'=0.170$.
}\label{fig9}
\end{figure}

As before, singular gauge transformation (\ref{n1}) renders the Hamiltonian periodic and we can solve it in momentum space using standard band structure techniques. The band structure for a square vortex lattice and a generic Fermi energy $\varepsilon_F$ is displayed in Fig.\ \ref{fig8}. The Majorana band shows a weak dispersion and is, once again, well described by the effective Majorana tight binding model discussed in Sec.\ II.E. One can extract the tunneling amplitudes $t$ and $t'$; these are plotted in Fig.\ \ref{fig9}. We find that they are reasonably well described by a simple heuristic formula
\begin{equation}\label{w14}
t\simeq A e^{-b k_F R}\left| \sin{ (k_F R)}\right|,
\end{equation} 
and a similar expression for $t'$ with parameters $A'$ and $b'$.
\begin{figure}[t]
\includegraphics[width = 8.0cm]{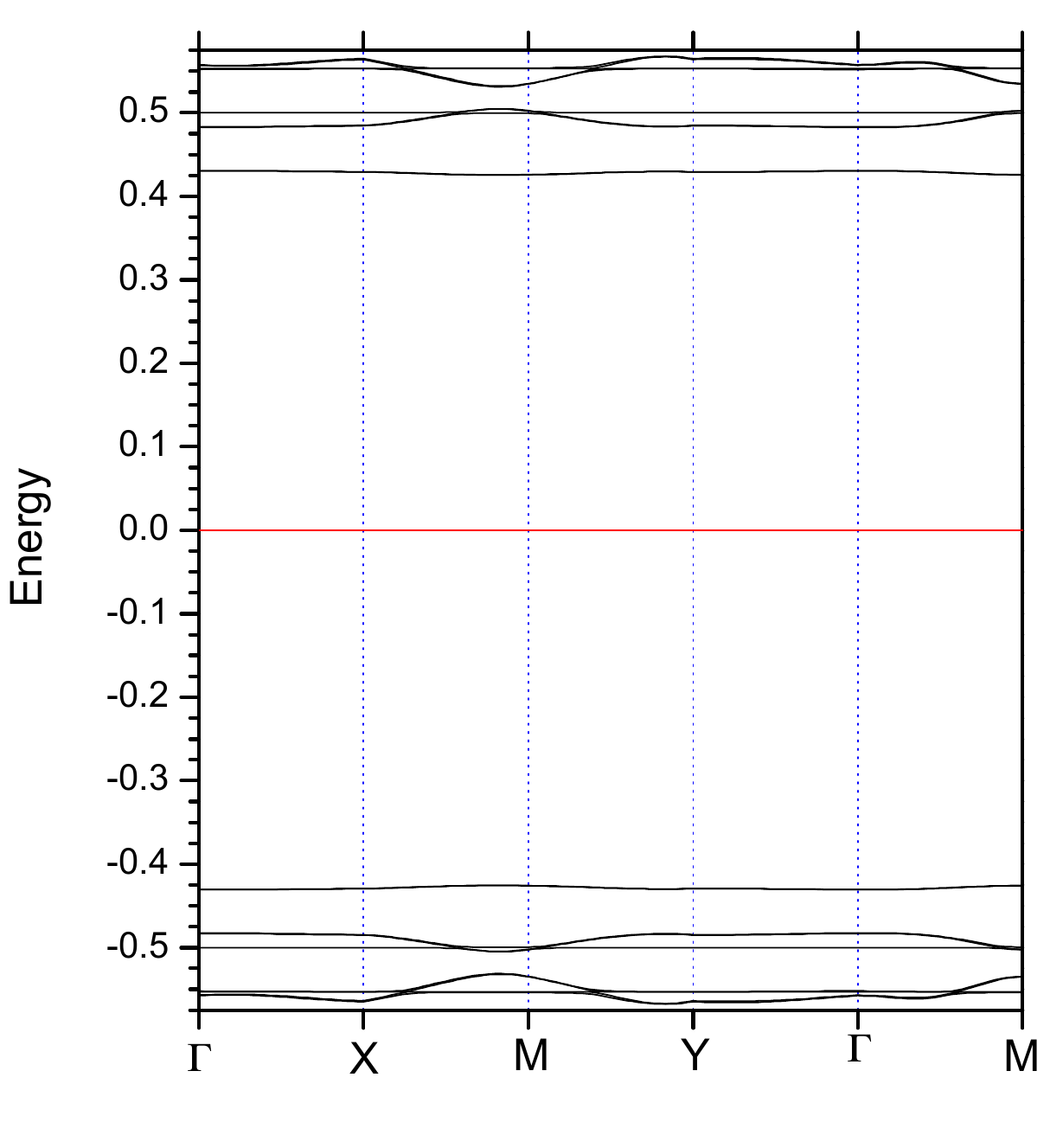}
\caption{Band structure for the lattice version of the Fu-Kane model with a deformed square vortex lattice, as explained in the text.  A $30\times 30$ magnetic unit cell is used with vortices located at (5,5) and (-5,-5) basis vectors. Also $\tau=0.5$,  $\Delta_0=0.4$, $m=0.5$ and $\varepsilon_F=0.0$.   
}\label{fig10}
\end{figure}

When $\varepsilon_F$ is tuned to the neutrality point we observe that both $t$ and $t'$ vanish which results in a completely flat Majorana band, as expected in the presence of the extra chiral symmetry discussed in Sec.\ II.B. While the tunneling amplitudes also vanish for certain nonzero values of $\varepsilon_F$, these are accidental zeros. Importantly, $\varepsilon_F=0$ is the only value for which $t$ and $t'$ (and presumably all other amplitudes) vanish simultaneously. To test that the flat band is indeed protected by the chiral symmetry (and not by some symmetry of the square vortex lattice) we performed simulations for a ``deformed'' square lattice. It is defined as follows: we keep the unit cell the same but within the unit cell we gradually move the B vortex closer to the A vortex along the diagonal line that connects them. When $\varepsilon_F=0$ the Majorana band remains completely flat for each A-B vortex separation, all the way to the point when the two vortices merge and  form a square lattice of doubly quantized vortices. This is illustrated in Fig.\ \ref{fig10}. 

i
\section{Conclusions}

Majorana zero modes bound to vortices in topological superconductors
form bands in the presence of a vortex lattice. We have demonstrated
that such bands are well described by simple tight binding models Eq.\
(\ref{kin1}) describing short ranged tunneling events between the
adjacent sites on the lattice. An interesting feature of these models
is the underlying non-trivial Z$_2$ gauge structure that is mandated
by the canonical anticommutation relations for the self-adjoint MZM
operators (\ref{can2}). We found that when the magnetic field
necessary for the vortex lattice formation is properly included the
Z$_2$ gauge factors obey the Grosfeld-Stern rule (\ref{kin2}),
previously derived in the context of MZMs in the Moore-Read fractional
quantum Hall state 
as well as vortex lattices
\cite{lahtinen0,lahtinen1} in the Kitaev spin model on the honeycomb
lattice \cite{kitaev2}. 
The hopping amplitudes are found to retain the full
periodicity of the vortex lattice and do not show any anomalies
suggested by previous works that neglect the applied magnetic field
\cite{biswas1,kou1,kou2}.

For periodic vortex lattices, such as the square and the triangular lattice, the resulting low-energy theory is typically gapped and topologically nontrivial. The latter property follows from the non-zero gauge flux implied by the Grosfeld-Stern phase factors. These, in turn, originate from the structure of the individual MZM wavefunctions and their overlap integrals. An intuitive understanding of this structure can be obtained from the following simple argument. Because of the self-adjoint property (\ref{can2}) of the MZM operators and their fermionic anticommutation relations the hopping amplitude between two sites is necessarily imaginary. This corresponds to the Z$_2$ phase $\pm i$. Now the simplest closed path on the lattice involves 3 distinct sites. The total Z$_2$ phase accumulated along such path is $\pm i$ which corresponds to a non-zero enclosed Z$_2$ flux. It follows that, generically, Majorana fermions defined on a lattice move in the background of a non-vanishing Z$_2$ gauge flux. In analogy with the Haldane model \cite{haldane1} one then expects the system to exhibit a non-zero Chern number and, in the geometry with open boundaries, protected gapless edge modes. This indeed has been noted in previous theoretical studies \cite{kou1,kou2}.

In the Fu-Kane model an interesting situation arises near the so called neutrality point where an extra chiral symmetry exists. The latter mandates that all hopping amplitudes exactly vanish resulting in the MZM band that is completely flat. This expectation is indeed borne out by our analytical as well as numerical calculations. Such completely flat bands are then highly susceptible to the effects of interactions and disorder. Some of the interaction and disorder effects have been explored in recent works \cite{chiu1,chiu2,pikulin1,rahmani1,cobanera1,rahmani2} and found various interesting interacting phases of Majorana zero modes in one and two dimensions.  

In a homogeneous superconductor the vortex lattice is expected to be perfectly periodic \cite{abrikosov1,tinkham1} and our results then directly apply. Many clean superconductors indeed exhibit such perfectly periodic vortex lattices. In a disordered superconductor, however, vortex lattice itself may become disordered. Our method for calculating the full electronic structure relies on translational invariance and cannot be directly applied to such disordered vortex lattices. However, our results indicate that the effective Majorana tight-binding model Eq.\ (\ref{kin1}) provides a good description of the low-energy physics. One thus expects that Majorana degrees of freedom in a disordered vortex lattice will be well described by the same tight-binding model in which the overlap integrals $\bar{t}_{ij}$ acquire a random component. Randomness in this model has been extensively studied \cite{lahtinen2,kraus1,vasuda1,lauman1,lauman2} and we expect these results to directly transfer to the present problem of vortex lattices with randomness. In addition, in a 2D system individual vortices can undergo thermal or quantum fluctuations around their equilibrium positions. The fate of the energy bands in this situation is an interesting problem which we leave for future study.

Recently, individual vortices have been experimentally observed by
scanning tunneling microscopy (STM) in 2D heterostructures combining a
topological insulator Bi$_2$Te$_3$ and a conventional superconductor
NbSe$_2$ \cite{xu1}. Evidence for possible MZMs bound in the cores of
such vortices has also been reported \cite{xu2}. Although these
experiments were not performed in the parameter regime where the MZM
band formation could be directly observed these developments suggest
that the results obtained in the present work can be experimentally
tested in the near future.

After this work was submitted for publication we became aware of a
preprint \cite{murray1} that reports results on the band structure of
$p_x+ip_y$ superconductor with vortices in agreement with our results. The
preprint also studies in detail the topological phases of the Majorana
bands and finds results that support conjectures presented in this
section, in addition to many other interesting new results.

\section{Acknowledgments and a dedication}
The authors are indebted to R. Biswas, A. Melikyan, J.M. Murray and O. Vafek for useful discussions and correspondence. The work reported here was supported by NSERC and by CIfAR. M.F.\ gratefully acknowledges the Institute of Quantum Information and Matter at Caltech where this project was first conceived. The authors dedicate this work to the memory of Zlatko Tesanovic, a collaborator and a friend, who pioneered the use of singular gauge transformations in unconventional superconductors.



\end{document}